\documentclass[fleqn,usenatbib]{mnras}

\usepackage{newtxtext,newtxmath}

\usepackage[T1]{fontenc}

\DeclareRobustCommand{\VAN}[3]{#2}
\let\VANthebibliography\thebibliography
\def\thebibliography{\DeclareRobustCommand{\VAN}[3]{##3}\VANthebibliography}

\usepackage{graphicx}	
\usepackage{amsmath}	
\usepackage{CJKutf8} 

\title[Uncovering the LMC's orbital history through its dynamical impact]{Uncovering the first-infall history of the LMC through its dynamical impact in the Milky Way halo}

\author[Sheng et al.]{Yanjun Sheng\begin{CJK*}{UTF8}{gbsn} (盛衍钧)\end{CJK*},$^{1}$\thanks{E-mail: Yanjun.Sheng@anu.edu.au}
Yuan-Sen Ting\begin{CJK*}{UTF8}{gbsn} (丁源森)\end{CJK*},$^{1,2,3,4,5}$
Xiang-Xiang Xue\begin{CJK*}{UTF8}{gbsn} (薛香香)\end{CJK*},$^{6,9}$
\newauthor Jiang Chang\begin{CJK*}{UTF8}{gbsn} (常江)\end{CJK*},$^{7}$
and Hao Tian\begin{CJK*}{UTF8}{gbsn} (田浩)\end{CJK*}$^{8,9}$
\\
$^{1}$Research School of Astronomy $\&$ Astrophysics, Australian National University, Cotter Rd., Weston, ACT 2611, Australia\\
$^{2}$School of Computing, Australian National University, Acton, ACT 2601, Australia\\
$^{3}$Department of Astronomy, The Ohio State University, Columbus, OH 43210, USA\\
$^{4}$Center for Cosmology and AstroParticle Physics (CCAPP), The Ohio State University, Columbus, OH 43210, USA\\
$^{5}$Department of Physics, Faculty of Science, Universiti Malaya, 50603, Kuala Lumpur, Malaysia\\
$^{6}$CAS Key Laboratory of Optical Astronomy, National Astronomical Observatories, Chinese Academy of Sciences,
Beijing 100101, People’s Republic of China\\
$^{7}$Purple Mountain Observatory, Chinese Academy of Sciences, Nanjing 210023, People’s Republic of China\\
$^{8}$Key Laboratory of Space Astronomy and Technology, 
National Astronomical Observatories, 
Chinese Academy of Sciences, 
Beijing 100101, P.R. China\\
$^{9}$Institute for Frontiers in Astronomy and Astrophysics, Beijing Normal University, Beijing, 102206, People’s Republic of China
}

\date{Accepted XXX. Received YYY; in original form ZZZ}

\pubyear{2024}

\begin{document}
\label{firstpage}
\pagerange{\pageref{firstpage}--\pageref{lastpage}}
\maketitle

\begin{abstract}
The gravitational interactions between the LMC and the Milky Way cause dynamical perturbations in the MW halo, leading to biased distributions of stellar density and kinematics. We run 50 high-resolution N-body simulations exploring varying masses and halo shapes of the MW and LMC to study the evolution of LMC-induced perturbations. By measuring mean velocities of simulated halo stars, we identify a discontinuity between the first-infall and second-passage scenarios of the LMC’s orbital history. In the first-infall, the Galactocentric latitudinal velocity hovers around 16 km/s for stars at 50-100 kpc, while it subsides to about 8 km/s in the second-passage scenario. We demonstrate that, this reduced perturbation magnitude in the second-passage scenario is mainly due to the short dynamical times of the Galactic inner halo and the lower velocity of the LMC during its second infall into the MW. Using a subset of $\sim 1100$ RR Lyrae stars located in the outer halo (50 kpc $\leq R_{\mathrm{GC}}<$ 100 kpc) with precise distance estimates from Gaia, we find the mean latitudinal velocity ($v_{b}$) in the Galactocentric frame to be $\langle v_{b} \rangle=18.1 \pm 4.1$ km/s. The observation supports the first-infall scenario with a massive LMC ($\sim$ $2.1 \times 10^{11} \mathrm{M}_{\odot}$) at infall, an oblate MW halo with a virial mass $M_{200}<1.4\times10^{12}\mathrm{M}_{\odot}$ and a flattening parameter $q>0.7$. Our study indicates that LMC-induced kinematic disturbances can reveal its orbital history and key characteristics, as well as those of the MW. This approach shows promise in helping determine fundamental parameters of both galaxies.

\end{abstract}




\section{Introduction}
\label{sec:intro}

The Large Magellanic Cloud (LMC) is the most luminous and prominent satellite of the Milky Way (MW) which has recently passed the pericenter of its orbit at about 50 kpc away from the galactic center, and has a mass only 5–10 times smaller than that of the Milky Way \citep[e.g.,][]{2013ApJ...764..161K,2018ApJ...857..101P,2018MNRAS.473.1218L,2019MNRAS.487.2685E}. Although interactions between the orbiting satellite galaxies and the dark matter halo of their hosts have been studied analytically since \cite{1943ApJ....97..255C}, our understanding of the LMC's orbital history and its impact on the Milky Way dynamics has only began to evolve in the last decade thanks to development of numerical simulations \citep[e.g.][]{2010ApJ...721L..97B,2015ApJ...802..128G,2018MNRAS.473.1218L,2018MNRAS.481..286L,2019ApJ...884...51G,2021MNRAS.501.2279V}{}{} and several large-scale surveys such as Gaia \citep{2018A&A...616A..12G}, APOGEE \citep{2017AJ....154...94M}, SEGUE \citep{2009AJ....137.4377Y}, LAMOST \citep{2012RAA....12.1197C} and H3 \citep{2019ApJ...883..107C}.

Before the more precise space-based measurements of the LMC's velocity became available, it was believed that the LMC had completed many orbits around the MW over the past few Gyrs \citep[e.g.,][]{1976ApJ...203...72T,1980PASJ...32..581M, 1982MNRAS.198..707L,1994MNRAS.266..567G}. This idea was challenged when the proper motion (PM) of the LMC was measured by \citet{2006ApJ...638..772K} using two-epoch Hubble space telescope data. The LMC's tangential velocity at $367 \pm 31$ km/s favors the first-infall scenario of the LMC's orbital history in which the Magellanic Clouds are on their first orbit around the Milky Way and the LMC has only one previous pericenter passage \citep{2007ApJ...668..949B}.

However, there is an alternative possibility regarding the LMC's orbital history, which is called the second-passage scenario. In this case, the LMC has completed an orbit around the MW and it has another pericenter passage before the most recent one. Despite the increasingly better measured motion of the LMC (e.g., \cite{2013ApJ...764..161K}), the debate of the first or second-passage scenario remains unsettled partly because
the orbital properties (e.g., orbital period and pericentre, apocentre distances) of the LMC can be sensitive to the variations in the mass distribution of the MW. While Gaia has provided more precise constraints on the Galactic total mass and density distribution, uncertainties at the level of 20–30$\%$ still remain from several independent methods, for example the rotation velocities of the Milky Way \citep[e.g.,][]{1970ApJ...159..379R,2008ApJ...684.1143X,2020MNRAS.494.4291C}{}{}, the phase-space distribution function \citep[e.g.,][]{1987ApJ...320..493L,2019ApJ...875..159E,2021MNRAS.501.5964D}{}{}, the locations and dynamics of stellar streams \citep[e.g.,][]{2010ApJ...712..260K,2016ApJ...833...31B,2019MNRAS.486.2995M}, the timing argument \citep[e.g.,][]{2008MNRAS.384.1459L,2016MNRAS.456L..54P,2020ApJ...888..114Z}{}{}, and matching observed satellite galaxies to simulations \citep[e.g.,][]{2011ApJ...743...40B,2014MNRAS.445.2049C,2018ApJ...857...78P}{}{}), which yield estimates of the virial mass of the MW that cluster around $10^{12} \mathrm{M}_{\odot}$ (see \citet{2020SCPMA..6309801W} for a recent review).

Recent independent lines of inquiries have led to evidences favoring the first-infall scenario of the LMC, but some of these evidence are still confronted. For instance:
\begin{itemize}
    \item The LMC's star formation rate has been unusually low until a recent burst starting about 3–4 Gyr ago \citep[e.g.,][]{2009AJ....138.1243H,2014MNRAS.438.1067M,2022MNRAS.513L..40M}, which could have been triggered by the compression of gas when the LMC entered the MW, supporting the first-infall scenario of the LMC's orbital history. However, this phenomenon can also be attributed to interactions with the Small Magellanic Cloud (SMC), as a synchronised ignition of star formation is found in the star formation histories of both the LMC and SMC around the same time period \citep{2022MNRAS.513L..40M}.
    \item Unlike the Sagittarius dwarf galaxy, which has completed several orbits around the MW over the past 6 Gyr \citep{2017ApJ...836...92D,2018MNRAS.481..286L}, there is no detected large-scale stellar stream following the past trajectory of the LMC, lending support to the first-infall scenario. Nonetheless, although we haven't discovered any noticeable stellar stream behind the LMC, some recent studies \citep[e.g.,][]{2020ApJ...905L...3Z,2022MNRAS.514.1266P,2023ApJ...956..110C} have shown evidences for stellar counterparts in the trailing and leading arms of the gas stream of the LMC.
    \item The Small Magellanic Cloud (SMC) and other satellites of the LMC would have been tidally stripped by the Milky Way if the LMC had made other close pericenter passages around the Galaxy. However, it was shown in \cite{2023arXiv230604837V} that even if the LMC had another encounter with the Milky Way between 5 and 10 Gyr ago at a distance exceeding 100 kpc, it could still retain almost all of its existing satellites, including the SMC. 
\end{itemize}
These findings highlight the limitations of current approaches and underscore the need for alternative methods to constrain the LMC's orbital history.

The past orbit of the LMC might imprint different gravitational interactions between the LMC and the MW halo \citep{2009MNRAS.400.1247C,2021ApJ...919..109G,2023arXiv230604837V}. Notably, orbiting satellite galaxies, such as the LMC, can create density asymmetries in the dark matter and stellar halos of their host galaxy, which subsequently affect the kinematic signature of the halo stars \citep{2015ApJ...802..128G,2016MNRAS.457.2164O,2018MNRAS.473.1218L,2018MNRAS.481..286L,2019ApJ...884...51G,2020MNRAS.494L..11P,2021MNRAS.506.2677E,2021ApJ...919..109G,2023Galax..11...59V}. In the MW-LMC system, the LMC induces two major responses in the MW’s halo: (a) the local effect, dynamical friction (DF) wake of the LMC, which is an overdensity of stars following the past trajectory of the LMC \citep{1943ApJ....97..255C,2009MNRAS.400.1247C,2016MNRAS.457.2164O,2019ApJ...884...51G,2021ApJ...916...55T}, and (b) the global effect, characterized by asymmetries in both kinematics and density of the MW on a large scale, arises mainly from the offset in net motion of the MW's inner regions relative to its outer regions (i.e. the reflex motion of the MW) \citep{1995ApJ...455L..31W,2015ApJ...802..128G,2019ApJ...884...51G,2021MNRAS.506.2677E}. These dynamical effects in the MW halo contain valuable information about the orbital frequency and past trajectory of the LMC.  

Recently, such dynamical effect was measured in the data. For instance, \cite{2021MNRAS.506.2677E} presented the detection of the bulk motion of the Milky Way’s stellar halo. They compiled a sample of about 500 stars with radial velocities in the stellar halo where the galactocentric radii exceed 50 kpc. Their sample of stars consists mainly of K giants and BHB stars from the SEGUE survey \citep{2011ApJ...738...79X,2014ApJ...784..170X}, which distances can be measured to a precision of $\sim$ 10$\%$. They show that in the outer halo, the radial velocities of stars become positive in the northern hemisphere and become negative in the southern hemisphere. Such bulk motion signatures are consistent with the simulated first-infall scenario, where the MW and the LMC have masses of $0.8 \times 10^{12} \mathrm{M}_{\odot}$ and $1.5 \times 10^{11} \mathrm{M}_{\odot}$, respectively, and the LMC has just completed its first pericentric passage. 


Additionally, \cite{2021Natur.592..534C} found evidence of halo stars’ density asymmetries induced by the LMC. They selected 1301 K giants from Gaia and WISE at a galactocentric radii range from 60 to 100 kpc, which was found to display a density asymmetry close to the LMC-induced perturbations described in previous simulations \citep{2019ApJ...884...51G}. Similar to the findings in \cite{2019MNRAS.488L..47B}, they also identified a prominent overdensity region located along the past trajectory of the LMC, which serves as evidence that the LMC is on its first orbit around the MW. However, the agreement between observation and simulation is not perfect. Another possible explanation to these overdensities was proposed by \cite{2023ApJ...951...26C}. They analysed their spectroscopic sample of distant halo stars (red giants) in the same overdense regions, and suggested that they may be due to debris from the Gaia-Sausage-Enceladus, an early massive merger in the Milky Way's history.

Tailored N-body simulations of the first-infall scenario have played a crucial role in interpreting observational features that are possibly induced by the LMC. However, the implications of the second-passage scenario have not been much studied through simulations, which inspired this work. The exploration of the MW-LMC interactions has largely been limited to a fiducial limited range in terms of the shape and the mass of the MW halo. To close this gap, in this study, we will present a comprehensive suite of simulations which consider the variation of different model parameters and include both scenarios where the LMC has made one or two pericenter passages. We also build upon previous studies by comparing the kinematic features of the latest sample of RR Lyrae stars with our simulations. 

This paper is organised as follows. In Section \ref{sec:data}, we describe the RR Lyrae star catalog that we used in this study. In Section \ref{sec:simulations}, we introduce the numerical simulations. In Section \ref{sec:analysis}, we analyse the dynamical effects induced by the LMC under different orbital histories (i.e. the first-infall and second-passage scenarios), and show how their mean values evolve with key parameters of MW-LMC system. We then compare the mean values of our RR Lyrae sample with that of the mock star catalog generated from simulations. We discuss the alignment of our findings with other observations, future implications and limitations of this study in Section \ref{sec:discussion} and conclude in Section \ref{sec:conclusion}.

\section{Data}
\label{sec:data}

\begin{figure}
    \includegraphics[width=\columnwidth]{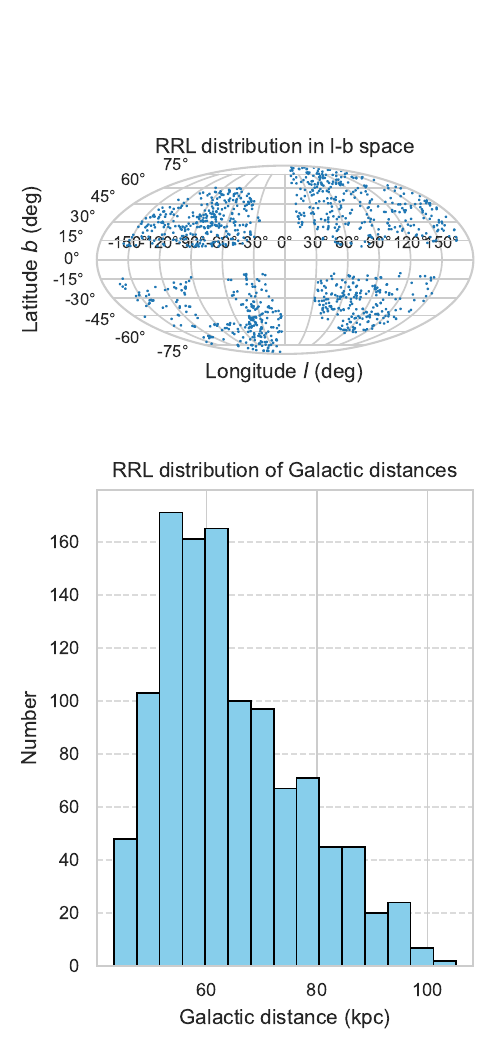}

    \caption{Our final 1,126 RR Lyrae stars. The on-sky spatial distribution in the Galactic longitude-latitude ($l$-$b$) space in shown in the top panel, and the distribution of the Galactic  distance is shown in the botton panel. The observation sample with which we will compare with our simulations includes RR Lyrae stars within the Galactocentric radius range of 50 kpc < $R_{\mathrm{gal}}$ < 100 kpc. The sample has removed notable halo substructures like LMC, SMC, Sagittarius stream ($\left|b_{\text{Sgr}}\right|<15^{\circ}$), stars distributed near the galactic disk $\left(|b|<10^{\circ}\right)$ and 4 visible dSph clusters to better study the dynamical effect of the LMC on the Milky Way halo.}
    \label{fig:RRL_sky_plot}
\end{figure}

We select stars from a Gaia-based catalog provided by \cite{2023ApJ...944...88L} (hereafter L23). This catalog contains 135,873 RR Lyrae stars (115,410 type RRab and 20,463 type RRc stars), with precise distance estimates (typical distance errors are only $3-4 \%$). RR Lyrae stars are core helium burning stars located on the horizontal branch;
RR Lyrae stars are standard candles due to their relatively uniform absolute magnitude ($M_{\mathrm{V}}$ $\sim$ 0.65 mag; e.g., \citealt{2008ApJ...676L.135C}), which allows them to be used as reliable distance indicators and traced to large distances. To derive precise distances from RR Lyrae stars, L23 cross-matched 135,873 RRLs which have their pulsation parameters and the Fourier decomposition parameters ($\phi_{31}$, $R_{21}$) from Gaia G-band light curves with a 2687 spectroscopic RRL sample to calibrate the period-pulsation parameter-metallicity relations. These calibrated relations, along with the G-band absolute magnitude-metallicity relations developed using $\sim 1000$ local bright local RRLs with accurate Gaia parallax measurements, are applied to all the RRLs and then used to derive distances. We focus on stars with Galactocentric radii larger than 50 kpc where the dynamical effect induced by the LMC should be prominent \citep{2019MNRAS.487.2685E,2020MNRAS.494L..11P,2023Galax..11...59V}. As the number of observed stars decreases rapidly at greater radii, we limit the sample to 100 kpc, to simplify the selection function. This truncation leave us with $15,594$ RR Lyrae.

We cross-match our RR Lyrae sample with Gaia DR3 to obtain the dynamical information (e.g., proper motion in the RA or DEC direction and their uncertainties) for these stars. Stars with $E(B-V)>0.3$ are masked from the catalog, as such high reddening values indicate significant interstellar dust and gas, which can affect the observed properties of stars (e.g., brightness and colors) and lead to potentially biased distance measurements. The values of $E(B-V)$, which are proportional to the extinction corrections, are adopted from \cite{1998ApJ...500..525S} for most regions of sky. However, for the low-latitude region with $|b| \leq 25^{\circ}$, the values are adopted from \cite{2023A&A...674A..18C}. For the LMC and SMC regions, the values are adopted from \cite{2021ApJS..252...23S}. For member RRLs of globular clusters, the extinction values are adopted from \cite{2010arXiv1012.3224H}. For the same reason of reducing the impact of extinction, we exclude stars within 10 degrees of the Galactic plane ($|b| > 10^{\circ}$), focusing our analysis on regions where the signals of halo stars are most prominent. 

To ensure a relatively clear signal from the smooth halo, we exclude known halo substructures from our sample, as they may represent unmixed debris from various accreted and tidally disrupted galaxies. The most prominent structure across the sky is the LMC and SMC, and we remove them in the $l$-$b$ space of the Galactic coordinate system. Member stars of the LMC and SMC are identified by requiring that they lie within a radius of $20^{\circ}$ and $10^{\circ}$ from the centers of the two systems, respectively. The positions of the centers of the LMC and SMC are $(l_{\text{LMC}} , b_{\text{LMC}} ) = (280.46^{\circ}, -32.75^{\circ} )$, as defined by \cite{2001AJ....122.1807V}, and $(l_{\text{SMC}} , b_{\text{SMC}}) = (302.95^{\circ}, -43.98^{\circ})$, as defined by \cite{2000A&A...358L...9C}. In addition, unmixed debris from the Sagittarius stream may confound the signal embedded in the smooth halo, so we mask out stars at $\left|b_{\text{Sgr}}\right|<15^{\circ}$ \citep{2021Natur.592..534C}, where $b_{\text{Sgr}}$ is the latitude in the frame of the Sagittarius orbital plane \citep{2014MNRAS.437..116B}. 

Moreover, we remove some visible clusters by applying DBSCAN (Density-Based Spatial Clustering of Applications with Noise) algorithm on the spatial coordinates of stars in the $l$-$b$ space. DBSCAN is an unsupervised machine learning algorithm used for clustering data points which finds core sample of high density and then perform a friends-of-friends search \citep{1982ApJ...257..423H}. We set the parameter \textbf{eps}, which defines how close points need to be to each other to be considered part of the same cluster, to 1.5, and \textbf{min-samples}, which defines the minimum number of points required to form a core point, to 10. Empirically, we found these parameters to be robust, and they identified 4 regions with known overdensities: Dra dSph, Ursa Minor dSph, Sextans dSph and Sculptor dSph. 

Our final sample consists of 1,126 RR Lyrae stars. This is a significant reduction from the initial $15, 594$ because the dominant sample of RR Lyrae stars is from the LMC and SMC, which is excluded in our case.
The spatial distribution as well as the distance of our RR Lyrae sample is shown in Figure ~\ref{fig:RRL_sky_plot}.

\section{Simulations}
\label{sec:simulations}

To investigate the influence of the LMC on the Milky Way and to validate our results derived from the RR Lyrae sample, we employ N-body simulations to model the interactions between the MW and LMC.
While hydrodynamical simulations {\color{blue} \citep{2017MNRAS.467..179G, 2023ApJS..265...44W}} include baryonic processes such as gas physics, star formation and stellar feedback would be ideal, for the purpose of this study, we contend that dark matter-only N-body simulations are sufficient for studying the dynamical influence of the LMC on the MW halo.

In Section~\ref{sec:Initial models of galaxies}, we will discuss the construction of the initial condition of the N-body models (the initial conditions) of the two galaxies, using analytical density distributions or potential models that are consistent with our current observational or theoretical constraints. The fidelity of these N-body models is crucial as they set the stage for the entire simulation, affecting the reliability of the final results. We generate the initial conditions for our N-body simulations using \textsc{galic} (GALaxy Initial Conditions) \citep{2014MNRAS.444...62Y}

We then model the MW-LMC interaction begins with low-resolution ($\sim 10^{6}$ particles) test simulations to determine the initial phase-space coordinates of the LMC, as discussed in Section~\ref{sec:orbital reconstruction}. This process requires iterative refinement of the LMC's initial phase-space coordinates to ensure that the present-day coordinates of its orbit in the simulation match those in the observation. And finally, in Section \ref{sec:high-resolution}, we run final high-resolution simulations ($\sim 10^{8}$ particles) to simulate the LMC-induced perturbations.

\renewcommand{\arraystretch}{1.3}
\setlength{\tabcolsep}{4.pt}
\begin{table*}
    \centering
    \caption{The present-day phase-space coordinates of the LMC in Galactocentric cartesian frame. All quantities are calculated from the measured values in \protect\cite{2013ApJ...764..161K}, by adopting the Solar reflex motion $V_{c, \text{ peak }}(8.29 \mathrm{kpc}) \approx 239$ km/s \protect\citep{2011MNRAS.414.2446M} and solar peculiar velocity $(U, V, W)_{\odot}=\left(11.1_{-0.75}^{+0.69}, 12.24_{-0.47}^{+0.47}, 7.25_{-0.36}^{+0.37}\right)$ km/s \protect\citep{2010MNRAS.403.1829S}. The standard errors on each component are calculated by Monte-Carlo sampling the errors on the measured proper motions\protect\footnotemark, radial velocities, and distances.}

	\label{tab:LMC_table}
	\begin{tabular}{cccccc} 
		\hline
		X (kpc) & Y (kpc) & Z (kpc) & $V_{x}$ (km/s) & $V_{y}$ (km/s) & $V_{z}$ (km/s)\\
		\hline
		$-1.06 \pm 0.33$ & $-41.05 \pm 1.89$ & $-27.83 \pm 1.28$ & $-57.60 \pm 7.99$ & $-225.96 \pm 12.60$ & $221.16 \pm 16.68$\\
		\hline
	\end{tabular}
\end{table*}


\subsection{Initial models of galaxies}
\label{sec:Initial models of galaxies}

The initial N-body systems of the MW and LMC are constructed with input density distributions (potential models) and velocity anisotropy using the \textsc{galic} code. We model the Milky Way with a potential similar to the MWPotential2014 from \cite{2015ApJS..216...29B}. This model consists of a NFW dark matter halo \citep{1997ApJ...490..493N} with a scale radius of $r_s = 16$ kpc. We explore a range of virial masses ($M_{200}$) of the halo, from $0.5\times10^{12} \mathrm{M}_{\odot}$ to $1.8\times10^{12} \mathrm{M}_{\odot}$, a conservative span that is consistent with estimates from other studies \citep{2020SCPMA..6309801W,2022ApJ...925....1S}. The virial mass and virial radius ($r_{200}$) are defined by the condition that the mean density $3M_{\mathrm{200}} /(4 \pi r_{\mathrm{200}}^3)$ is 200 times the critical density of the universe. Furthermore, we consider oblate, spherical and prolate halo shapes with various flattening parameters (q, indicating flattening perpendicular to the Galactic disk), taking values of 0.5, 0.7, 0.9, 1.0, 1.1 and 1.3. In addition to halos with an isotropic velocity profile ($\beta(r)=0$), we also investigate halos with a radially varying anisotropy velocity profile \citep{2006NewA...11..333H}: $\beta(r)=-0.15-0.2 \alpha(r) ; \alpha(r)=\frac{\mathrm{d} \ln \rho(r)}{d \ln r}$. 

Apart from the halo, to ensure that the gravitational potential is realistic in the inner regions of the MW (< 30 kpc) and to accurately track the center of mass of the system, we include a Miyamoto Nagai disk \citep{1975PASJ...27..533M}, with the mass $M_{d}=6.8\times10^{10} \mathrm{M}_{\odot}$, the disk scale height $b = 0.28$ kpc, the scale length $a = 3$ kpc, and a Herquist bulge \citep{1990ApJ...356..359H},with the mass $M_{b}=0.5\times10^{10} \mathrm{M}_{\odot}$, the scale radius $a_{h}=0.54$ kpc. Since the combined mass of the disk and bulge is subtantially smaller than that of the halo, we assume all MW models have the same disk and bulge parameters as MWPotential2014, regardless of the (varied) halo parameters. 

Other parameters of the Milky Way model are derived from the virial mass $M_{\mathrm{200}}$ of the halo, with the flattening parameter q factored in through the spheroid radius $r \equiv \sqrt{ X^2+Y^2+(Z / q)^2}$. The virial radius $r_{200}$, is calculated through the equation 
\begin{equation}
r_{\mathrm{200}}=206 h^{-1} \mathrm{kpc}\left(\frac{200}{97.2}\right)^{-1 / 3}\left(\frac{M_{\mathrm{200}}}{10^{12} h^{-1} \mathrm{M}_{\odot}}\right)^{1 / 3}
\end{equation}
where $h=1.0$, as the Hubble constant is $100 \mathrm{~km} \mathrm{~s}^{-1} \mathrm{Mpc}^{-1}$ in the internal units of \textsc{galic}. Therefore, the concentration parameter $c$ of the halo is $c=r_{200}/r_{s}$, where $r_{s}$ is fixed at 16 kpc. Unlike MWPotential2014, the MW halo is represented by a Hernquist profile in \textsc{galic}, so we calculate the scale radius $a_{h}$ in order to make sure that the enclosed mass at the virial radius equals to that of the NFW profile: 
\begin{equation}
a_h=\frac{r_{200}}{c} \sqrt{2\left[\ln (1+c)-\frac{c}{1+c}\right]}
\end{equation}
Note that the input $r_{200}$ and $c$ are properties not only of the halo but also of the entire Milky Way galaxy, accounting for the mass of the disk and bulge \citep{2022ApJ...927..131B}. To simultaneously satisfy these two conditions, we increase the value of $r_{200}$ by changing the $M_{\mathrm{200}}$ of the halo to the total mass of the MW $M_{\mathrm{total}}$. We then use the calculated scale radius $a_{h}$ and the updated $r_{\mathrm{200,new}}$ to derive a new concentration parameter $c_{\mathrm{new}}$. Both $r_{\mathrm{200,new}}$ and $c_{\mathrm{new}}$ serve as our final input parameters to \textsc{galic}.

As for the model of LMC, we only vary the virial mass ($M_{200}$) of its halo. The LMC is modeled using a spherical Hernquist profile to represent its DM halo. Although the actual LMC has a stellar disk, we do not include this structure in simulations because we are interested in the impact of the LMC
on the MW halo kinematics, and dominant perturbations come from the DM halo of the LMC \citep{2002AJ....124.2639V}. The Hernquist scale radius $a_{h}$ is constrained by the fact that the circular velocity of the halo at a distance of 8.7 kpc from the center of the galaxy is about 92 km/s \citep{2014ApJ...781..121V} and through the equation
\begin{equation}
v_c(r=8.7\mathrm{kpc})=\frac{\sqrt{G M_{200} r}}{r+a_h}\approx 92 \mathrm{km/s}
\end{equation}
In this study, we construct 8 LMC models with virial masses specifically set at $0.7, 0.9, 1.1, 1.3, 1.5, 1.7, 1.9$ and $2.1 \times 10^{11} \mathrm{M}_{\odot}$. 

To study the LMC-induced kinematic perturbations in the MW halo, it is essential that the N-body models of both galaxies have achieved stability before their interaction is simulated in \textsc{gadget-4} \citep{2021MNRAS.506.2871S}. The stability of an N-body galaxy system is demonstrated by the convergence of the halo's velocity anisotropy profile $\beta(r)$ and density profile $\rho(r)$ in the simulation to their analytical form. When running the initial models, we follow the results from the convergence test by \cite{2019ApJ...884...51G}, where they evolve different MW N-body models in isolation over 5 Gyr, and show that after 2.5 Gyr, the variations in $\beta(r)$ and $\rho(r)$ is minimal. To be conservative, we evolve the N-body models of the LMC and the MW separately in isolation (without the other galaxy) for a duration of 3 Gyr before we integrating them into a single snapshot and simulating their interaction starting from the moment the LMC first crosses the virial radius of the Galactic halo. 

\footnotetext{Since the proper motion measurements of the LMC do not include covariances, here we only compare the reconstructed coordinates from N-body simulations with the marginal distributions of individual phase-space measurements in the Galactocentric cartesian frame.}

\subsection{Orbital reconstruction}
\label{sec:orbital reconstruction}
We note that the initial condition construction as described above only allows to initiate the N-body particles of the LMC and the MW. However, to properly combine the two galaxies into a snapshot requires us to calculate the relative position of the two galaxies while satisfying the current observation of the LMC motion with respect to the MW, given the (varied) parameters. And for that, we need to first reconstruct the LMC's past orbit around the MW for the  different given parameters. 

As an approximation, given that we have the current position and velocity of the LMC relative to the MW from observations, as well as the analytical potential models of the MW and LMC that were previously used to generate initial N-body systems, we can integrate the orbit backwards in time through the following ODE system to get the initial 6D phase-space coordinates of the LMC when it first crosses the virial radius of the MW:  
\begin{equation}
\ddot{\boldsymbol{x}}_{\mathrm{LMC}}=\frac{-d \Phi_{\mathrm{bulge}, \mathrm{MW}}}{d \boldsymbol{x}}+\frac{-d \Phi_{\mathrm{disk}, \mathrm{MW}}}{d \boldsymbol{x}}+\frac{-d \Phi_{\mathrm{halo}, \mathrm{MW}}}{d \boldsymbol{x}}+\frac{\boldsymbol{F}_{\mathrm{df}}}{\boldsymbol{M}_{\mathrm{LMC}}}
\end{equation}
\begin{equation}
\ddot{\boldsymbol{x}}_{\mathrm{MW}}=\frac{-d \Phi_{\mathrm{LMC}}}{d \boldsymbol{x}}
\end{equation}

These equations describe how the center position of both galaxies evolve under their mutual gravitational force, as well as the dynamical friction that arises from the gravitational interactions between a massive moving object and the surrounding particles, resulting in an asymmetrical pull that slows down the object. Here, $\boldsymbol{x}$ represents the position vector between the center of two galaxies, $\Phi_{\mathrm{MW}}$ and $\Phi_{\mathrm{LMC}}$ represent the static potential of each galaxy centered at $\boldsymbol{x}_{\mathrm{MW}}$ and $\boldsymbol{x}_{\mathrm{LMC}}$, and $\boldsymbol{F}_{\mathrm{df}}$ is the dynamic friction. 

We perform the integration using \textsc{galpy} \citep{2015ApJS..216...29B}, which is a Python package for galactic dynamics and supports orbit integration in a variety of potentials. The current coordinates and associated errors of the LMC used in this work are given in Table ~\ref{tab:LMC_table}. These values are derived following the same method as \cite{2020ApJ...893..121P}, and are calculated using measurement of observed sky position $(\text{RA},\text{DEC})=(78.76 \pm 0.52, -69.19 \pm 0.25)$ deg, distance $r=49.6 \pm 2.3$ kpc, line-of-sight velocity $v_{\mathrm{los}}=64 \pm 7$ $\mathrm{km} \mathrm{s}^{-1}$, and proper motion $(\mu_{W}, \mu_{N})=(-1.910 \pm 0.020, 0.229 \pm 0.047)$ $\operatorname{mas} \mathrm{yr}^{-1}$ from \cite{2013ApJ...764..161K}.

However, the approximations employed in orbit integration calculations inevitably introduce systematic errors in the derived initial coordinates of the LMC. These errors prevent the precise reproduction of the LMC's present-day position and velocity in simulations. For instance, considering that the mass of the LMC is only 5 to 10 times smaller than that of the Milky Way, the LMC's gravitational influence causes the MW’s disk to be displaced from the Galactic center of mass. This results in a time-varying potential field for the MW, while our analytical models assume constant potentials for both galaxies during integration. In addition, accurately predicting the value of dynamical friction applied to the LMC remains a challenge. The classical Chandrasekhar theory \citep{1943ApJ....97..255C} only offers a reasonable approximation for the order of magnitude of the frictional force. In galpy, the dynamical friction exerted on a satellite galaxy with mass $M_{\mathrm{sat}}$ at position $\boldsymbol{x}$ moving with a velocity $\boldsymbol{v}$ through a background halo density $\rho(\boldsymbol{x})$ adopts the following form:
\begin{equation}
\boldsymbol{F}_{\mathrm{df}}=-2 \pi G^2 M_{\mathrm{sat}}^2 \ln (1+\Lambda^2) \rho(\boldsymbol{x})\left[\operatorname{erf}(X)-\frac{2 X}{\sqrt{\pi}} \exp \left(-X^2\right)\right] \frac{\boldsymbol{v}}{v^3}
\end{equation}
Here, $X=v /\sqrt{2\sigma}$ where $\sigma$ is is the one-dimensional galaxy velocity dispersion. The factor $\Lambda=\frac{r/\gamma}{\max \left(r_{\mathrm{hm}}, G M_{\text{sat}} /v^2\right)}$, where $r_{\mathrm{hm}}$ is the half-mass radius of the satellite and $\gamma=1$. 

To determine the appropriate initial coordinate of the LMC that ensures alignment between the simulated and observed present-day coordinates, we employ an iterative adjustment process. Starting from the initial coordinate obtained through numerical integration (which we consider as the first guess for the iteration), we apply the Gauss-Newton iteration method (detailed in Appendix \ref{sec: Gauss-Newton}). This involves conducting a series of low-resolution simulations (with particle masses of $1 \times 10^6 \mathrm{M}_{\odot}$ and $\sim 10^{6}$ particles), and iteratively refining the initial coordinate until the final state of the N-body orbits closely matches the LMC's observed present-day position and velocity. Our choice of particle number is consistent with some recent merger simulations \citep[e.g.,][]{2018MNRAS.481..286L,2019ApJ...884...51G,2021ApJ...923...92N}{}{}, which is sufficient to accurately model the gravitational interaction between two galaxies while facilitating the iteration process. As a result, our N-body simulations reproduce the LMC’s present-day position and velocity within 2$\sigma$ of the observed values. Further details on the number of simulations required and the CPU hours utilized are provided in Appendix \ref{sec: Gauss-Newton}.

\subsection{Reveal the LMC-induced perturbations with high-resolution simulations}
\label{sec:high-resolution}
After we find the most satisfactory orbital initial coordinates of the LMC, we run this suit with higher resolution to reveal the LMC-induced dynamical effects in the halo. As is discussed in \cite{2007MNRAS.375..425W}, the ability of a N-body simulation to capture the small scale (several kpc) perturbations mainly depends on the number of particles it used. The particle mass of both MW and LMC models in our final high-resolution simulations is $1 \times 10^4 \mathrm{M}_{\odot}$ with a total number of particles of $\sim10^{8}$. Previous studies \citep[e.g.,][]{2009MNRAS.400.1247C,2019ApJ...884...51G,2023arXiv230604837V}{}{} have tested that this amount of particles is sufficient for our research purposes. We also performed a convergence test in Appendix \ref{sec: Convergence test}.


In this study, we assume softening lengths for all kinds of particles in \textsc{gadget-4} to be $\epsilon=600-700(60-70)$ pc for the low (high) resolution simulations, following the criteria of \cite{2003MNRAS.338...14P}, $\epsilon=\frac{4 r_{200}}{\sqrt{N_{200}}}$, where $N_{200}$ is the total number of particles within $r_{200}$ of the MW. For the first-infall scenario, we run the MW-LMC interactions for 2 Gyr in \textsc{gadget-4}, with each simulation taking approximately 6,000 CPU hours. While for the second-passage scenario, we run the simulations for 4-5 Gyr, each consuming approximately 20,000 CPU hours. 

\section{Results of simulations}
\label{sec:analysis}

\begin{figure}
	\includegraphics[width=\columnwidth]{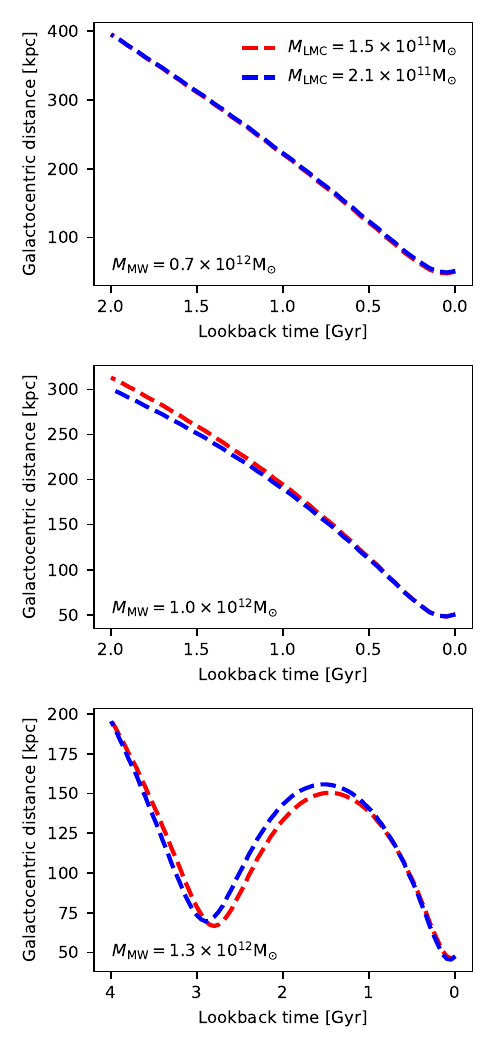}
    \caption{The distance between the centers of the LMC and the Milky Way as a function of lookback time, tracing back to the moment when the LMC first enters the virial radius of the MW. Each panel in the figure corresponds to a representative MW model with a spherical halo and a virial mass of $0.7\times10^{12}\mathrm{M}_{\odot}$ (top), $1.0\times10^{12}\mathrm{M}_{\odot}$ (middle), and $1.3\times10^{12}\mathrm{M}_{\odot}$ (bottom). For each MW model, we show the simulated trajectories of two LMC models with different virial masses at the time of infall: $M_{\text{LMC}}=1.5\times10^{11}\mathrm{M}_{\odot}$ (red) and $M_{\text{LMC}}=2.1\times10^{11}\mathrm{M}_{\odot}$ (blue). For MW masses of $M_{\text{MW}}=0.7$ and $1.0\times10^{12}\mathrm{M}_{\odot}$, the LMC's orbital history corresponds to the first-infall scenario. However, a 30$\%$ increase in the virial mass of the MW halo can cause a shift in the LMC's orbital history, transitioning it to the second-passage scenario.}
    \label{fig:LMC_orbits}
\end{figure}

\begin{figure}
	\includegraphics[width=\columnwidth]{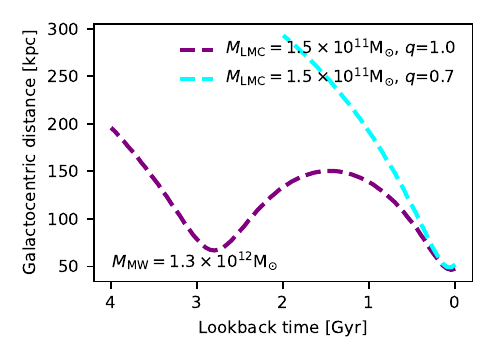}
    \caption{The shape of the Milky Way plays a crucial role in determining the LMC's orbital history. We demonstrate the past trajectories of the fiducial LMC model ($M_{\text{LMC}}=1.5\times10^{12}\mathrm{M}_{\odot}$) around spherical ($q$ = 1.0) and oblate ($q$ = 0.7) MW halos, respectively. While the LMC follows a second-passage orbital scenario around the spherical halo, it exhibits a higher orbital energy around an oblate halo. This results in the LMC being less gravitationally bound to the Milky Way, leading to the first-infall scenario.}
    \label{fig:LMC_orbits_q}
\end{figure}

We now analyze the results of our simulations in this section. We first present the orbital histories of the LMC in a few representative cases which guide our selection of model parameters for further investigation. Subsequently, we conduct a comparative analysis of the LMC-induced dynamical effects in both the first-infall and second-passage scenarios. We provide an explanation for the reduced perturbation magnitudes observed in the second-passage scenario. We then quantify the magnitudes of the dynamical effects with their average values and explore how they are influenced by the key model parameters identified earlier. Finally, we compare the perturbation magnitudes in the latitudinal velocity of halo stars, located within a Galactocentric distance range of 50 to 100 kpc, between our RR Lyrae star sample and the simulated star samples derived from our full simulation grid. Our findings suggest that the magnitude of these perturbations can serve not only as an indicator of the LMC's orbital history but also as a mean to place constraints on the properties of the halos of both the MW and the LMC. In Section \ref{sec: LMC orbital history}, we show the past trajectories of the LMC in the N-body simulation in three representative cases. In Section \ref{sec:dynamical effects}, we first reproduce some of the known LMC-induced dynamical effects in the MW halo for the first-infall scenario, and then explore them in the second-passage scenario. In Section \ref{sec:statistical properties}, we quantify the kinematic perturbations of halo stars with summary statistics, their mean values, and illustrate the reason for our choice. In Section \ref{sec:assessment of features}, we evaluate the influence of some key parameters of the MW and the LMC on the magnitude of the kinematic perturbations in Galactocentric frame, starting from our fiducial MW-LMC model. Finally, in Section \ref{sec: compare obs sim}, we compare the mean latitudinal velocity in the Galactocentric frame between the RR Lyrae sample and the simulated star catalogs derived from our grid of MW-LMC simulations. We then use the results of the comparison to constrain the orbital history of the LMC and some key properties of the two galaxies.

\subsection{Orbital histories of the LMC}
\label{sec: LMC orbital history}

The LMC's orbital history is affected by the mass distribution of both the LMC and the Milky Way. As the mass of the MW increases, the gravitational interaction between the two galaxies becomes stronger, which can shift the LMC's orbital trajectory from a first-infall to a second-passage scenario. While changes in the LMC's mass do not directly trigger a shift in the orbital scenario, they are significant in determining the LMC's initial kinematic energy at the time of its infall. 

Figure \ref{fig:LMC_orbits} shows the simulated trajectory of the LMC in three representative cases, each characterised by a spherical MW halo with unique halo mass ($M_{\text{MW}}=0.7,1.0,1.3\times10^{12}\mathrm{M}_{\odot}$ and $q$ = 1.0, represented in the top, middle and bottom panel respectively). For each MW halo mass, we consider two LMC masses ($M_{\text{LMC}}=1.5,2.1\times10^{11}\mathrm{M}_{\odot}$, illustrated by the red and blue dashed lines respectively). The $x$-axis represents the lookback time, traced back to the moment when the LMC first enters the virial radius of the MW, while the $y$-axis indicates the Galactocentric distance of the LMC. The LMC’s orbital history shown in the upper two panels corresponds to the first-infall scenario where the LMC enters into the MW for the first time about 2 Gyr ago. In the bottom panel, with a 30$\%$ increase in the mass of the Milky Way's halo, the orbit transitions to the second-passage scenario and the time of the LMC's first infall extends to 4 Gyr ago. 

The properties of these orbits are consistent with those described in \cite{2023Galax..11...59V}. Given that the LMC's present-day phase space coordinates are kept fixed, more massive LMCs are associated with lower initial kinematic energy when the MW has a relatively small virial mass, while this relationship is reversed as the MW mass becomes larger. This phenomenon results from the interplay of two opposite effects. (1) The dynamical friction becomes larger for the more massive LMC and induces higher energy cost, which implies that it should have a higher initial kinematic energy to reach its current phase-space position. (2) The reflex motion of the MW becomes larger when the LMC is more massive, so in order to maintain the same relative velocities around the MW, the LMC needs to start with a lower kinematic energy. The second factor outweighs the first one when the MW has a lower virial mass, but more massive MWs tend to have smaller reflex motion and the effect of dynamical friction becomes more important. 

Empirical evidence suggests that the shape of the galactic halo, characterized by its flattening parameter $q$, is not necessarily spherical \citep[e.g.,][]{2013ApJ...773L...4V,2018ApJ...859...31H,2021MNRAS.508.5468H,2004MNRAS.351..643H,2016MNRAS.460..329B}{}{}. Therefore, a natural question arises: how does the variation in $q$ affect the transition between the first-infall and second-passage scenarios of the LMC? In Figure \ref{fig:LMC_orbits_q}, we show the LMCs’ orbital histories in an oblate MW halo, which has the same mass of galaxies ($M_{\text{MW}}=1.3\times10^{12}\mathrm{M}_{\odot}$, $M_{\text{LMC}}=1.5\times10^{11}\mathrm{M}_{\odot}$) as the model in the bottom panel of Figure \ref{fig:LMC_orbits} but the MW halo is flattened in the direction perpendicular to the Galactic disk with a flattening parameter $q$ = 0.7. Due to the flattened shape of the halo, the mass of the Milky Way enclosed within a given radius (e.g., the inner 100 kpc) decreases by about 30\%. As a result, the Galactic potential at the present-day location of the LMC has increased relative to that in a spherical halo. Since the current kinematic energy of the LMC relative to the MW remains constant in both cases, this change in potential energy leads to a higher orbital energy, making the LMC less gravitationally bound to the Milky Way. Thus, compared to the case with a spherical ($q$ = 1.0) halo of equal mass, where the LMC makes two pericentric passages, the LMC’s past orbit around an oblate halo belongs to the first-infall scenario this time. This underscores the significance of treating the shape of the MW halo as a crucial variable in our simulations.

\begin{figure*}
	\includegraphics[width=2.0\columnwidth]{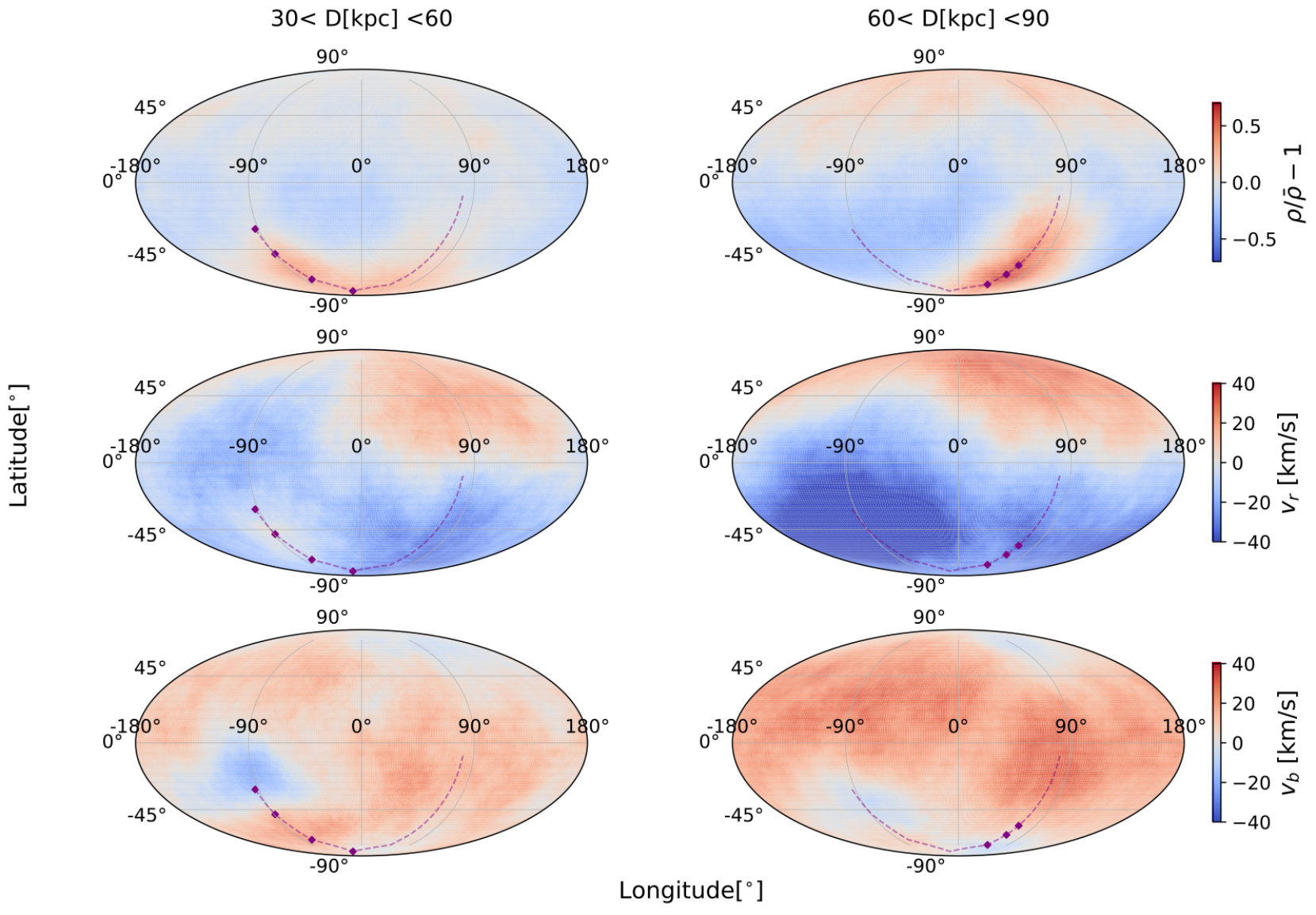}
    \caption{We present the dynamical effects induced by the LMC on the density, radial velocity ($v_{r}$) and latitudinal velocity ($v_{b}$) fields of the MW halo in all-sky maps at two distinct Galactocentric radii intervals: 30 kpc to 60 kpc and 60 kpc to 90 kpc. These maps are generated from the present-day snapshot of a representative MW-LMC simulation, where the MW model has a spherical halo with a virial mass of $M_{\text{MW}}=0.7\times10^{12}\mathrm{M}_{\odot}$, a spherical halo ($q=1.0$) and an isotropic velocity profile ($\beta(r)=0$). The LMC model in this simulation has a virial mass of $M_{\text{LMC}}=1.5\times10^{11}\mathrm{M}_{\odot}$. In these maps, the trajectory of the LMC is marked with a dashed purple line, with solid diamonds highlighting instances when the LMC was within $30-60$ kpc or $60-90$ kpc of the Galactic disk. 
    The top row illustrates density fluctuations in the stellar halo, with the color bar reflecting the relative variations between the local density (computed from the nearest 1000 particles at each grid point) at a given point in the sky and the spherical average density. The middle and bottom rows show the radial and latitudinal velocities, respectively, with their color bars representing the local mean values.}
    \label{fig:0.7_map}
\end{figure*}

\begin{figure*}
	\includegraphics[width=2.0\columnwidth]{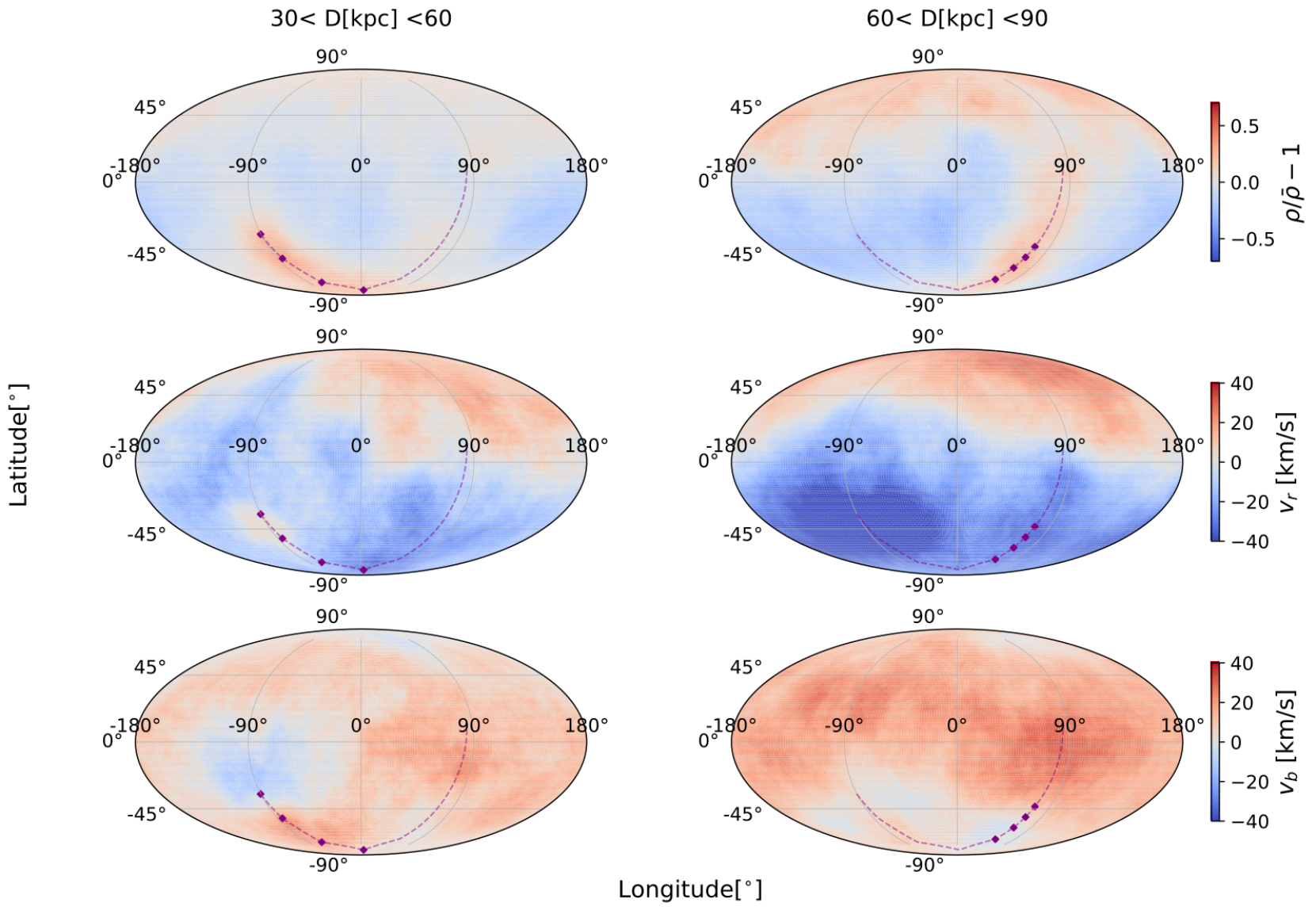}
    \caption{Similar to Figure \ref{fig:0.7_map}, but here we assume the virial halo mass of the MW in the simulation to be $1.0\times10^{12}\mathrm{M}_{\odot}$. This simulation remains to be one of the first passage scenarios. Compared to Figure \ref{fig:0.7_map}, while the kinematic signatures (middle and bottom rows) remain largely consistent, there is a noticeable shift in the position of the dynamical friction overdensity along the orbit of the LMC.}
    \label{fig:1.0_map}
\end{figure*}

\begin{figure*}
	\includegraphics[width=2.0\columnwidth]{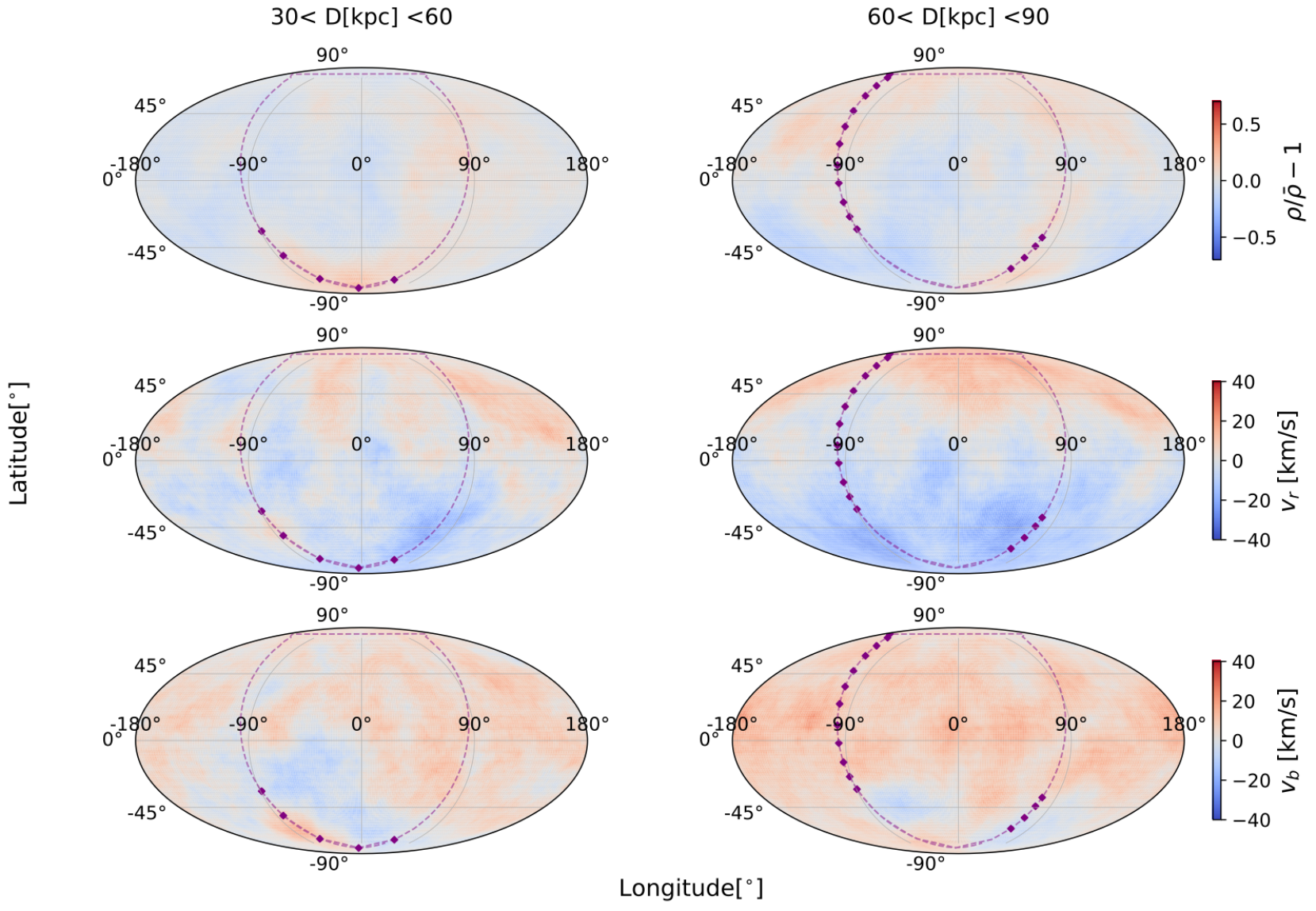}
    \caption{Similar to Figure \ref{fig:0.7_map} and Figure \ref{fig:1.0_map}, but with the MW halo mass increased to $1.3\times10^{12}\mathrm{M}_{\odot}$. Due to the higher mass of the MW halo, the LMC's past orbit now includes two pericentric passages, resulting in a second-passage orbital scenario. In this case, while we can still identify density and kinematic features associated with local and global effects, the magnitudes of these perturbations are subtantially reduced. This is due to the short dynamical time (less than 0.3 Gyr) in the Galactic inner halo (within the inner 30 kpc). The present-day dynamical effects depicted in this figure are primarily a result of the LMC's second infall, where the LMC had lower kinematic energy compared to the first-infall scenario.}
    \label{fig:1.3_map}
\end{figure*}

\begin{figure*}
	\includegraphics[width=2.0\columnwidth, height=0.3\textwidth]{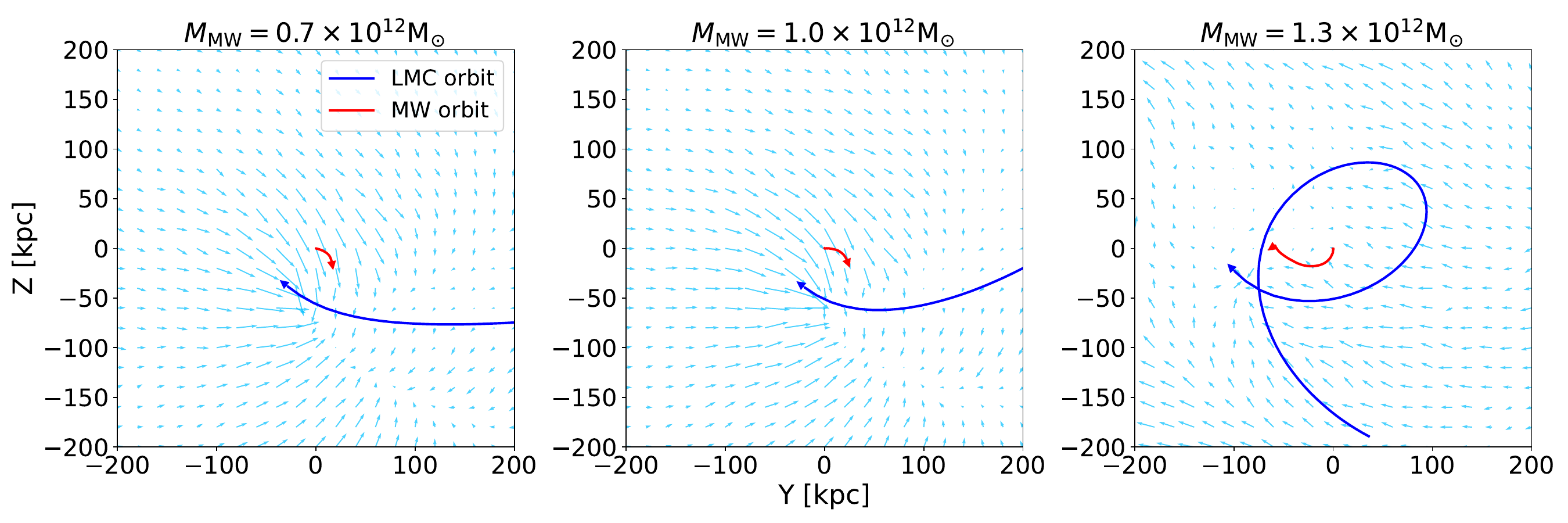}
    \caption{Dynamical effects in three representative cases (using the same MW-LMC systems as in Figure \ref{fig:0.7_map}-\ref{fig:1.3_map}) viewed from the orbital plane of the LMC. All the figures are shown in the $Y$-$Z$ plane of the present-day snapshots, and the thickness of the slice equals to 10 kpc. The light blue vectors in the background represent the mean velocities of halo stars, with the value of each grid is computed over the nearest 1000 neighbors. The blue curve and red curve represent the orbit of the LMC and MW respectively. In the first-infall scenario (left and middle column), the LMC enters the MW halo from the right side of the Galactic disk, passes below it and then moves towards the upper left. The inner halo of the MW experiences a consistent acceleration towards the LMC, ultimately attaining a relatively high velocity displacement relative to the outer halo. However, in the second-passage scenario (right column), the LMC enters the MW halo from a location below the disk and then orbits it in a clockwise direction. The final velocity of the MW's inner halo is largely due to the acceleration induced by the LMC's second pericentric passage, resulting in a much smaller displacement relative to the outer halo.}
    \label{fig:velocity_plot}
\end{figure*}

\subsection{Dynamical effects in the Milky Way halo}
\label{sec:dynamical effects}

Having established how the halo properties of the two galaxies can affect the LMC's past trajectory, we now turn our attention to the dynamical effects induced by the LMC in the Milky Way halo. We first aim to reproduce some of the known density and kinematic features for the first-infall scenario. This will demonstrate the robustness of our simulations and set the stage for our subsequent exploration of the second-passage scenario, which was largely unexplored. 

We show the Galactocentric sky maps of density and velocity perturbations in three representative cases in Figure \ref{fig:0.7_map}-\ref{fig:1.3_map}, using the same simulation suits described in Figure \ref{fig:LMC_orbits} ($M_{\text{MW}}=0.7,1.0,1.3\times10^{12}\mathrm{M}_{\odot}$, $M_{\text{LMC}}=1.5\times10^{11}\mathrm{M}_{\odot}$, $q$ = 1.0 and velocity profile of the MW halo is isotropic $\beta(r)=0$). We show the density contrast (top row, the relative variations between the local density and the average density), the line-of-sight velocity (middle row) and the latitudinal velocity (bottom row) within the stellar halo at two radii intervals: 30-60 kpc (left column) and 60-90 kpc (right column). Maps of the inner 30 kpc of the halo are not shown in this study, as there are no significant perturbations in this region because the majority of particles within this area moving at a similar velocity to that of the Galactic disk. We note that the density and kinematic features of the first-infall scenario demonstrated in Figure \ref{fig:0.7_map} and Figure \ref{fig:1.0_map} are consistent with those obtained from first-infall simulations in previous studies \citep{2019ApJ...884...51G,2020MNRAS.498.5574E,2023arXiv230604837V}. The dynamical influence of the LMC on the MW can be broadly understood as follow:

\textbf{Local effect:} As the LMC attracts nearby stars (represented by particles in the simulation), it accumulates them along its trajectory, leading to a local stellar overdensity in the density map (top row) that traces the LMC's historical orbit.

\textbf{Global effect:} There is a a disparity in dynamical times between the inner and outer halo (with the dynamical time of the MW halo at 30, 50, 100 kpc being approximately 0.3, 0.5, and 1.0 Gyr, respectively). The shorter dynamical times in the inner halo cause stars in the inner galaxy to react faster to the passage of the LMC, while stars in the outer halo are less sensitive to the LMC's trajectory. This differential response results in a relative displacement between the two regions. The inner region of the MW is accelerated downward (below the Galactic disk) towards the LMC while the outer region has negligible changes in net velocity and therefore appears moving upward in the Galactocentric frame. The net upward motion of the outer halo generates an overdensity in the north and an underdensity in the south (top row). This motion also give rise to a similar north-south dipole asymmetry in the radial velocity (middle row), with stars in the northern sky are moving away from us and southern stars approaching us and a positive bias in the latitudinal velocity (bottom row). As for the second-passage scenario, as shown in Figure \ref{fig:1.3_map}, we can still identify density and kinematic features associated with local and global effects, but their magnitudes are considerably reduced.

\begin{figure*}
	\includegraphics[width=2.0\columnwidth]{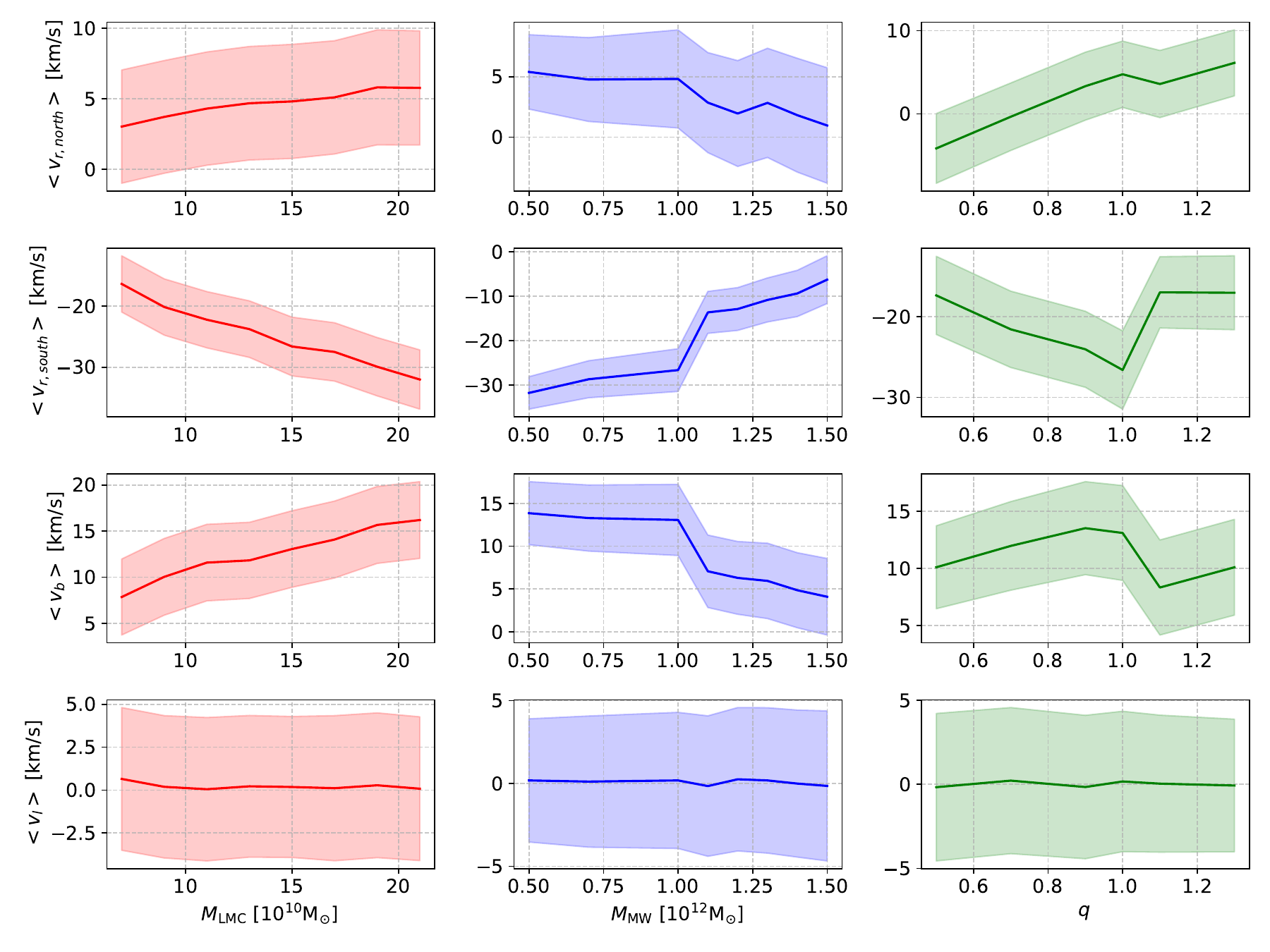}
    \caption{Variations in the average values of stellar radial ($v_{r}$), latitudinal ($v_{b}$), and longitudinal ($v_{l}$) velocities, as observed in the Galactocentric frame, correlate with some key parameters of the MW-LMC system. We assume a fiducial MW-LMC system, characterized by parameters $M_{\text{MW}}=1.0 \times 10^{12} \mathrm{M}_{\odot}$, $M_{\text{LMC}}=1.5 \times 10^{11} \mathrm{M}_{\odot}$ and $q$ = 1.0, and proceed to investigate the neighboring parameter space by changing one parameter at a time while keeping the others fixed. The mean velocities and their uncertainties are derived using the bootstrap technique, with each bootstrap sample matching the same size and the spatial distribution as the RR Lyrae sample. In the left column, models with a consistent MW potential but varying LMC mass reveal a proportionality between the magnitude of perturbations and the mass of the LMC. In the middle column, models with identical LMC mass but different MW potentials demonstrate that there is a transition in the LMC's orbital history from the first-infall to the second-passage scenarios when the halo mass of the MW exceeds $1.0\times10^{12}\mathrm{M}_{\odot}$. In the right column, models with the same MW and LMC mass but different flattening parameters $q$ show that for a halo mass of $1.0\times10^{12}\mathrm{M}_{\odot}$, the LMC's orbital history shifts from the first-infall scenario to the second-passage scenario as the flattening parameter changes from $q$ $\leq1.0$ to $q$ = 1.1.}
    \label{fig:stat_mean}
\end{figure*}

\textbf{Within the Galactocentric distance range of 30 to 60 kpc:} The mean density contrast (i.e. $\rho / \bar{\rho}-1$, representing the relative fluctuations between the local density and the average density across the sky) in regions of the dynamical friction wake decreases from approximately 0.3-0.5 in the first-infall scenario (Figures \ref{fig:0.7_map} and \ref{fig:1.0_map}) to about 0.1 in the second-passage scenario (Figure \ref{fig:1.3_map}). In the northern hemisphere, the mean radial velocities remain close to zero across both scenarios. In the southern hemisphere, the mean radial velocities drop from about -15 km/s in the first-infall to -6 km/s in the second-passage. The mean latitudinal velocities decrease from around 8 km/s in the first-infall to approximately 4 km/s in the second-passage.

\textbf{Within the Galactocentric distance range of 60 to 90 kpc:} The mean density contrast in regions of the dynamical friction wake remains similar to the previous distance range. In the northern hemisphere, the mean radial velocities are consistent at about 3-4 km/s in both scenarios. In the southern hemisphere, the mean radial velocities decrease from around -30 km/s in the first-infall to about -10 km/s in the second-passage. The mean latitudinal velocities drop from approximately 15 km/s in the first-infall to about 6 km/s in the second-passage.

This phenomenon is due to the short dynamical time (less than 0.3 Gyr) in the Galactic center region (inner 30 kpc). Following the LMC's first pericenter passage and its subsequent move to the apocenter position in its orbit, the inner galaxy attained a new equilibrium with its surroundings, leading to an almost negligible relative displacement between the inner and outer halo. Consequently, the local and global dynamical effects observed in Figure \ref{fig:1.3_map} predominantly stem from the LMC's second infall, during which it has a lower relative velocity with respect to the MW compared to the first-infall scenario in Figure \ref{fig:0.7_map} and Figure \ref{fig:1.0_map}. In order to maintain the present-day relative velocity around the MW, the Galactic inner halo experienced a smaller acceleration this time, resulting in weaker global dynamical effects. We show the analysis of this result in detail in Section \ref{sec: low_second_passage}.

To provide a clearer perspective on our analysis, we also present the LMC-induced kinematic perturbations from an alternative point of view in Figure \ref{fig:velocity_plot}. Here, the reflex motions of the Milky Way in three representative cases are projected onto the $y$-$z$ plane of the present-day simulation snapshots, which is centered at ($x$, $y$, $z$) = (0, 0, 0), with the $y$ and $z$ axis ranging from -200 kpc to 200 kpc. This plane is closely align with the LMC's orbital plane. We show the trajectories of the Galactic disk (red curve) and the LMC (blue curve) in the inertial frame of the simulation. The light blue vectors in the background of each plot represent the mean velocities of halo stars, with the value of each grid is computed over the nearest 1000 neighbors. In the first-infall scenario (left and middle column), the LMC begins its journey from the right side of the Galactic disk ($y$ = 200 kpc), passes below it and then moves towards the upper left. Meanwhile, as the LMC approaches, the Galactic disk starts its motion from the center of the plot and moves towards the position of the LMC.

In terms of halo stars, we find that while stars in the inner halo moves towards the current position of the LMC with approximately the same velocities as the Galactic disk, those in the outer halo are less accelerated. Regions through which the LMC passes (on the lower right side of the disk) are first pulled towards the LMC as it approaches from the right, and then pulled in the opposite direction towards the Galactic disk as the LMC moves left, thus the net velocity of them are close to zero. In contrast, regions where the LMC is moving towards (on the left side of the disk) experience consistent acceleration towards the LMC and ultimately attain relatively high velocities. Additionally, stars in the inner halo of the Milky Way move towards the LMC along with the disk, while stars in the outer halo do not experience a comparable amount of downward acceleration due to their larger dynamical times. As a result, when observed from the disk, the net velocity of the outer halo appears to move upward and exhibits a inward movement in the southern hemisphere and an outward movement in the northern hemisphere of the sky along the line-of-sight direction.

However, the LMC and the Galactic disk follows a more complex trajectory in the second-passage scenario (right column). The LMC starts from a location well below the Galactic plane ($z$ = -190 kpc), orbiting the disk in a clockwise direction. As it ascends in an elliptical orbit, the LMC passes through its first pericenter, rises above the Galactic plane, and then reaches the apocenter to the upper right of the disk. Subsequently, the LMC starts the return arc of its orbit, swinging back toward the lower left side of the disk and passing through its second pericenter during this phase. Meanwhile, the disk's motion is primarily influenced by the LMC's two pericentric passages. It originates from the center of the plot and shifts leftward towards the LMC's position. As the LMC advances further towards its apocenter, the velocity vector of the disk remains relatively unchanged. 

However, the disk experiences a noticeable acceleration towards the lower left as the LMC passes through its second pericenter. The motion of halo stars is notably more complicated for the second passage scenario. As previously mentioned, due to the short dynamical times of stars in the inner halo, stars in the inner halo reached equilibrium with the outer halo and there was little remaining dynamical signals induced by the LMC's first pericenter passage when the LMC started its second infall with reduced kinematic energy. Thus, the final displacement of the Galactic center is primarily attributed to the acceleration resulting from the LMC's second pericenter passage, and the final velocities of the inner Galaxy are subtantially lower in comparison to the first-infall scenario.

\subsection{Measure the kinematic perturbations with mean values}
\label{sec:statistical properties}

Having observed qualitative differences in the LMC-induced dynamical effects between the first-infall and second-passage scenario, we now aim to quantify the perturbations in the radial and latitudinal velocities of halo stars (the bottom two rows of Figures \ref{fig:0.7_map}-\ref{fig:1.3_map}) and use these information to constrain the orbital history of the LMC. In our analysis, we will focus on the simple summary statistics of perturbations in stars' velocity field, primarily their average values. Due to the large measurement uncertainties in the proper motion of remote halo stars ($\sim$0.2 mas/yr for an RR Lyrae at a distance of 50 kpc, corresponding to the velocity uncertainty of 100 km/s), the average values can help mitigate the large uncertainties associated with the individual star’s proper motion measurements, while reflecting the magnitude of the LMC-induced global effects. 

In addition, we chose to focus on the velocity field, because precisely measuring LMC-induced density asymmetries is complicated by the differences between the distribution functions of the simulated dark matter halo and our tracer of interest (i.e. RR Lyrae stars in this study), with the inference of the tracer's distribution relies heavily on our precise understanding of the survey's selection function. However, unlike the density field, the velocity field are more dependent on the overall gravitational potential and less directly dependent on the precise density distribution of the stellar halo.  

We are also aware of other methods have been proposed in previous studies to quantitatively study the LMC-induced perturbations in density and velocity fields. For example, \cite{2020ApJ...898....4C} and \cite{2021ApJ...919..109G} investigated the spherical harmonic expansion of the velocity field and the basis function expansion of the density field within the simulated MW stellar halos derived from \cite{2019ApJ...884...51G}. While these methods show the potential for disentangling the dynamical effects from the initial structure of the MW halo, the method might face more complications for our study due to the measurement uncertainties and limited sky coverage. The expansion of the density field can also be subject to the complex selection function that is beyond the scope of this study. For these reasons, we keep using simpler summary statistics and defer the discussion of more detailed distributions of density and kinematic perturbations to future studies.

\subsection{Quantitative assessment of the kinematic features}
\label{sec:assessment of features}
In this section, we aim to assess the influence of some key parameters of the MW-LMC system on the magnitude of the LMC-induced perturbations in the velocity field. We begin with the fiducial MW-LMC system, characterized by parameters $M_{\text{MW}}=1.0 \times 10^{12} \mathrm{M}_{\odot}$, $M_{\text{LMC}}=1.5 \times 10^{11} \mathrm{M}_{\odot}$ and $q$ = 1.0. We then explore the surrounding parameter space by sequentially changing one parameter at a time, while holding the others constant. Our study concentrates on the Galactocentric distance range that spans from 50 to 100 kpc (the same distance range as our RR Lyrae sample). This particular range was chosen because it not only well reflects the magnitude of the perturbations induced by the LMC, but also encompasses a relatively larger volume of observational data. 
 
Figure \ref{fig:stat_mean} illustrates the evolution of the mean values of stars' radial velocities in the northern and southern hemispheres (first and second rows), latitudinal velocities (third row), and longitudinal velocities (bottom row), with respect to the infall mass of the LMC (left column), the halo mass of the MW (middle column), and the flattening parameter of the MW halo (right column). We perturb the phase-space coordinate of our simulation particles with measurement uncertainties of 10$\%$ in distance measurements, 20 km/s in radial velocities and 100 km/s in tangential velocities. These estimates are based on the observational uncertainties of the 1,126 RR Lyrae stars in our dataset. The mean velocities and their uncertainties (which include both sampling noise and introduced measurement errors) are derived using the bootstrap technique by repeatedly drawing samples of particles with replacement from the simulation, with each bootstrap sample having the same size and the spatial distribution as the RR Lyrae sample. In this process, we exclude prominent substructures (i.e. LMC, SMC, Sagittarius stream, and the vicinity of the galactic disk) in the $l$-$b$ space as illustrated in the upper panel of Figure \ref{fig:RRL_sky_plot}, and then randomly select simulation particles to match the radial distribution shown in the lower panel of Figure \ref{fig:RRL_sky_plot}. Given the nearly uniform distribution of RR Lyrae stars across the sky, the differences in uniformity within the $l$-$b$ space between simulated and real stars should have minimal impact on their statistical properties.


\textbf{Varying mass of the LMC}: Models with the same MW potential parameters (i.e. MW mass, halo concentration and flattening parameter) but different LMC mass at infall are demonstrated in the left column, the magnitude of perturbations in radial ($v_{r}$) and latitudinal ($v_{b}$) velocities is proportional to the mass of the LMC. With the mass of the LMC evolves from $0.7\times 10^{11} \mathrm{M}_{\odot}$ to $2.1\times 10^{11} \mathrm{M}_{\odot}$, the magnitude of perturbations are almost doubled. For the mean radial velocity in the southern hemisphere, it changes from about -16 km/s to about -32 km/s. Similarly, for the mean latitudinal velocities, the value changes from about 8 km/s to about 16 km/s. This effect arises because the parameters of our MW model remain fixed, resulting in constant dynamical times for the MW's inner and outer halos. Consequently, a more massive LMC can induce a larger displacement in the net motion of the inner halo relative to the outer halo.

\begin{figure*}
	\includegraphics[width=2.0\columnwidth]{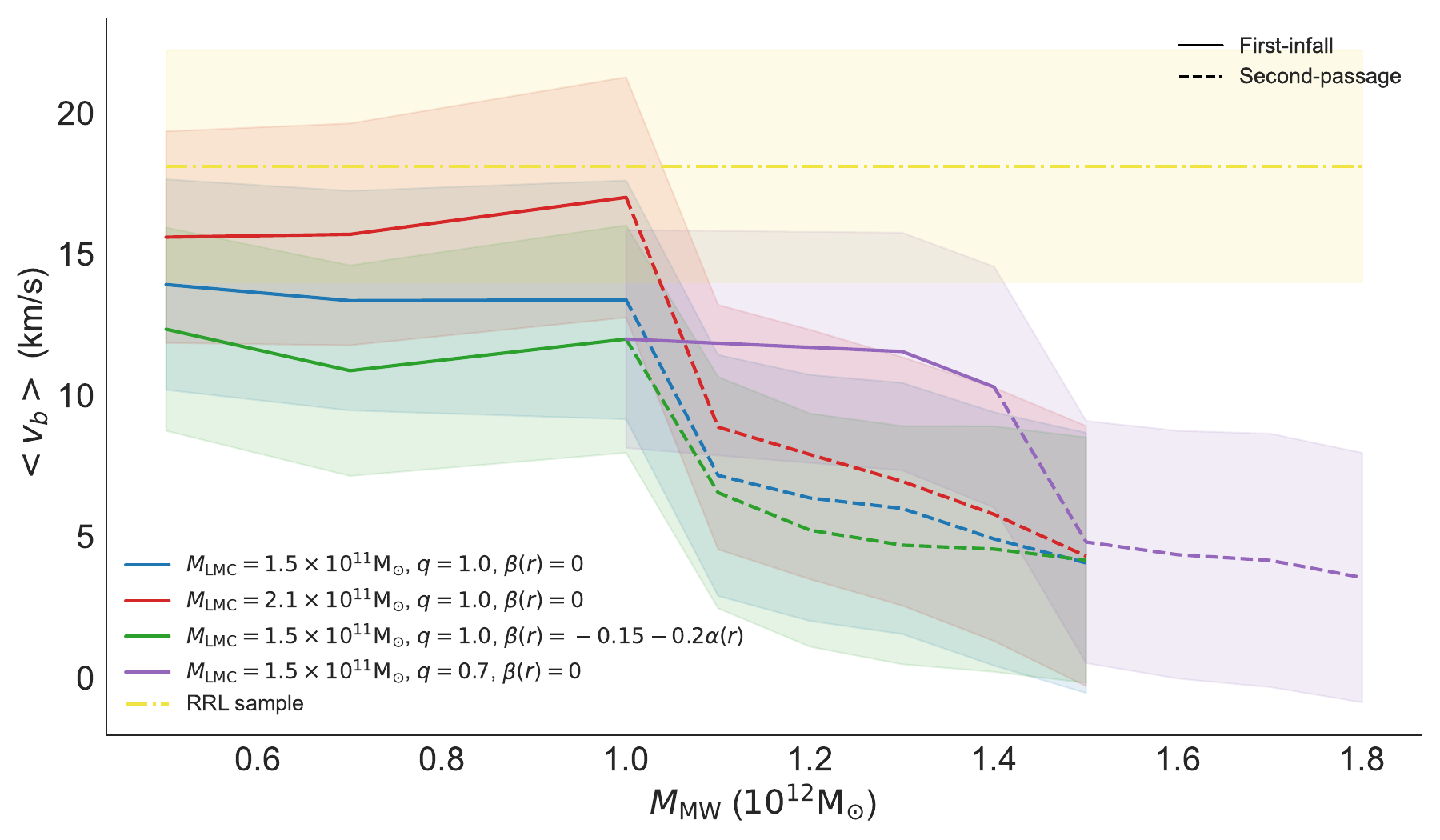}
    \caption{Comparison of the mean latitudinal velocity ($v_{b}$) as calculated in the Galactocentric frame between observational data and simulation results. The mean and uncertainty of the RR Lyrae stars are shown with the yellow dash-dotted line and the corresponding shaded region. For the mock data, their mean values and uncertainties are represented by the solid (first-infall), dashed (second-passage) curves and shaded regions in different colors. The blue shaded region represents MW-LMC systems with a LMC mass of $1.5\times10^{11} \mathrm{M}_{\odot}$, featuring a spherical Galactic halo with isotropic ($\beta(r)=0$) velocity profile but varying MW halo masses. The red shaded region represents the MW-LMC models with the same MW parameters as the blue shaded region but with a heavier LMC ($M_{\text{LMC}}=2.1\times10^{11} \mathrm{M}_{\odot}$). The green shaded region maintains the same MW and LMC potentials as the blue shaded region but assumes the MW halo has a radially anisotropic ($\beta(r)=-0.15-0.2 \alpha(r)$, where $\alpha(r)=\frac{\mathrm{d} \ln \rho(r)}{d \ln r}$) velocity profile. The purple shaded region represents the cases where the shape of the MW halo in the blue shaded region transitions to an oblate shape ($q$ = 0.7), with its minor axis perpendicular to the Galactic disk. We find that the RR Lyrae sample has a mean latitudinal velocity $\langle v_{b}\rangle=18.1\pm4.1$ km/s, which supports the first-infall scenario as the likely orbital history for the LMC. In addition, our findings favour a massive LMC model with virial mass close to $2.1\times10^{11}\mathrm{M}_{\odot}$ at the time of infall, an oblate MW halo with the flattening parameter $q$ larger than 0.7 and a mass of the MW halo less than $1.4\times10^{12}\mathrm{M}_{\odot}$ (for detailed discussion, see Section \ref{sec:mass-shape degeneracy}).}
    \label{fig:RRL_sim_compare}
\end{figure*}

\textbf{Varying mass of the MW halo}: On the other hand, models with a constant LMC mass but varying MW halo masses (middle column) demonstrate a shift in the LMC's orbital history from the first-infall scenario to the second-passage scenario when the MW mass exceeds $1.0\times10^{12}\mathrm{M}_{\odot}$. Following this transition, the magnitude of kinematic signals decreases by about a factor of two, resulting in a discontinuity between the first-infall and second-passage scenarios. The underlying cause of this transition is the increased mass enclosed within a given radius from the Galactic center, which leads to a reduced gravitational potential at the present-day location of the LMC in a more massive halo. Given that the LMC's current kinematic energy relative to the MW remains unchanged regardless of variations in halo mass, this decrease in potential energy results in a lower orbital energy and the LMC becomes more gravitationally bound to the Milky Way.

In addition, stars in more massive MW halos tend to be less perturbed by the LMC's infall. In the first-infall scenario, when the mass of the MW halo increases from $0.5\times10^{12}\mathrm{M}_{\odot}$ to $1.0\times10^{12}\mathrm{M}_{\odot}$, the magnitude of the radial velocities in southern hemisphere decreases by about 5 km/s, from -32 km/s to -27 km/s, and the magnitude of the latitudinal velocities decreases by about 1 km/s, from 14 km/s to 13 km/s. In the second-passage scenario, where the mass of the MW halo ranges from  $1.1\times10^{12}\mathrm{M}_{\odot}$ to $1.5\times10^{12}\mathrm{M}_{\odot}$, the inverse relationship between perturbation magnitudes and halo masses persists. The magnitude of radial velocities in the southern hemisphere decreases by about 8 km/s, from -14 km/s to -6 km/s, and the magnitude of latitudinal velocities decreases by about 3 km/s, from 7 km/s to 4 km/s. This phenomenon can be attributed to changes in the dynamical times of the MW's outer halo. As the MW's halo mass increases, the dynamical times of both the inner and outer halos decrease, with a more pronounced reduction in the outer halo. Consequently, stars in the outer halo are more quickly to reach equilibrium with those in the inner halo, leading to a smaller displacement in the net motion between the inner and outer halos.

\textbf{Varying flattening parameter of the MW halo}: Moreover, the orbital history of the LMC is also sensitive to the variation in the shape of the MW halo. Models with the same MW and LMC mass ($M_{\text{MW}}=1.0 \times 10^{12} \mathrm{M}_{\odot}$, $M_{\text{LMC}}=1.5 \times 10^{11} \mathrm{M}_{\odot}$) but different flattening parameters $q$ (ranging from 0.5 to 1.3, as shown in the right column) show a discontinuity in the magnitude of kinematic perturbations when transitioning from $q$ $\leq$ 1.0 to $q$ = 1.1. This transition suggests that, for this MW mass, the orbital history of the LMC aligns with the first-infall scenario when the MW halo is oblate or spherical. However, it can shift to the second-passage scenario when the MW halo becomes prolate. We present a discussion on the degeneracy between halo mass and shape when it comes to determining the orbital history of the LMC in section \ref{sec:mass-shape degeneracy}. Similar to the transition associated with variations in the MW halo mass, the enclosed mass within a given radius is lower in an oblate halo compared to a prolate halo. In our cases, the enclosed mass within the inner 100 kpc increases by 22\% when the flattening parameter changes from 0.9 to 1.1. Consequently, the gravitational potential at the LMC's present-day location is significantly higher in an oblate halo. Given that the LMC's current kinematic energy relative to the MW remains unchanged in both cases, this variation in potential energy results in the LMC having a higher orbital energy and being less gravitationally bound when orbiting an oblate halo, whereas it possesses a lower orbital energy and is more gravitationally bound when orbiting a prolate halo.

We also find an increase in the magnitude of perturbations when the flattening parameter of the halo increases in both orbital scenarios. In the first-infall scenario, when the flattening parameter of the MW halo increases from 0.5 to 1.0, the magnitude of the radial velocities in southern hemisphere increases by about 10 km/s, and the magnitude of the latitudinal velocities shows an increase of about 3 km/s. In the second-passage scenario, where the flattening of the MW halo ranges from 1.1 to 1.3, the magnitude of radial velocities in the southern hemisphere increases by about 1 km/s, and the magnitude of latitudinal velocities increases by about 2 km/s. This effect can once again be linked to changes in the dynamical times of the MW's outer halo, which are influenced by the shape of the halo. As the flattening parameter of the halo increases, the mean density within a certain radius decreases, leading to longer dynamical times for the outer halo. Consequently, stars in the outer halo take longer to reach equilibrium with those in the inner halo, resulting in a larger displacement in the net motion between the inner and outer halos.

\subsection{Compare observation with simulations}
\label{sec: compare obs sim}

We now proceed to compare the observed latitudinal velocities ($v_{b}$) from the RR Lyrae sample with the outcomes derived from our grid of MW-LMC simulations. Our previous analysis of the LMC-induced perturbations was carried out in the Galactocentric coordinate system within the context of N-body simulations, where the observer is placed at the center of the MW. This perspective provides us with a clearer understanding of the LMC's impact on Galactic halo and facilitates the identification of asymmetries in the density and velocity fields. Given that our RR Lyrae sample is limited to 5-dimensional phase-space coordinates (lacking the radial velocity data), we use \textbf{reflex-correct} function from \textsc{gala} package to correct for the solar reflex motion in the stars' proper motions and calculate their velocities in the Galactocentric frame. For this correction, we set the position and velocity of the Sun in the Galactocentric cartesian frame to (8.30, 0.00, 0.02) kpc \citep{2019MNRAS.482.1417B} and (11.10, 244.24, 7.25) km/s \citep{2010MNRAS.403.1829S,2017MNRAS.465...76M} respectively.  

Figure \ref{fig:RRL_sim_compare} demonstrates the mean latitudinal velocity $\langle v_{b}\rangle$ in the Galactocentric frame. The $x$-axis represents the virial mass of the MW halo, which spans from $0.5\times 10^{12}\mathrm{M}_{\odot}$ to $1.8 \times 10^{12}\mathrm{M}_{\odot}$, while the $y$-axis shows the average latitudinal velocity for both the RR Lyrae sample and the simulated star catalog. We estimate the uncertainty of RR Lyrae stars by Monte-Carlo sampling the errors in their observables (i.e. proper motions measurements from Gaia and distances estimates from \cite{2023ApJ...944...88L}) 10,000 times. For simplicity, we do not introduce measurement covariances as the distance and proper motion measurements in our study are done independently, and we deem the impact of the proper motion covariances on our analysis to be small. While for the statistical values in simulations, as elaborated in the Section \ref{sec:assessment of features}, we first introduce realistic measurement uncertainties in stars' distances and velocities to the present-day simulation snapshots, and then derive the mean velocities and their uncertainties using the bootstrap technique by repeatedly drawing samples of particles with replacement from the simulation, with each bootstrap sample having the same particle numbers and the spatial distribution as the 1,126 RR Lyrae sample. We find that the uncertainties in our simulated catalogs are larger than those in the observational data. This discrepancy arises because our estimates account for both the intrinsic noise introduced into the distances and velocities of particles, and additional statistical sampling noise generated when we randomly select particles to create bootstrap samples.

The blue shaded region in the figure represents MW-LMC systems with a LMC mass of $1.5\times10^{11} \mathrm{M}_{\odot}$ before it enters into the MW, a spherical Galactic halo with an isotropic velocity profile, but varying MW halo masses. The mean latitudinal velocities $\langle v_{b}\rangle$ for these models are about $13\pm4$ km/s in the first-infall scenario (represented by the solid line), where the MW halo mass ranges from $0.5\times10^{12} \mathrm{M}_{\odot}$ to $1.0\times10^{12} \mathrm{M}_{\odot}$, and decrease to below $7\pm4$ km/s in the second-passage scenario (represented by the dashed line), with the halo mass ranging from $1.1\times10^{12} \mathrm{M}_{\odot}$ to $1.5\times10^{12} \mathrm{M}_{\odot}$. The red shaded region represents models with the same MW potentials and velocity profile as the blue shaded region but with a heavier LMC mass at infall ($M_{\text{LMC}}=2.1\times10^{11} \mathrm{M}_{\odot}$). In these models, the mean latitudinal velocities increase by about 3 km/s to $16\pm4$ km/s in the first-infall scenario and by 1 km/s to below $8\pm4$ km/s in the second-passage scenario.

In addition to the models featuring a MW halo with isotropic velocity profile ($\beta(r)$ = 0), the green shaded region maintains the same MW and LMC potentials as the blue shaded region but assumes a MW halo with radially varying anisotropic velocity profile \citep{2006NewA...11..333H}: $\beta(r)=-0.15-0.2 \alpha(r)$, where $\beta(r)$ is defined as $\beta(r)=1-\frac{\sigma_t(r)^2}{2 \sigma_r(r)^2}$, with $\sigma_r$ and $\sigma_t$ being the radial and tangential velocity dispersion of halo stars, respectively, and $\alpha(r)=\frac{\mathrm{d} \ln \rho(r)}{d \ln r}$. This choice is motivated by previous findings indicating that a radially varying anisotropic velocity profile can alter the morphology of the LMC-induced dynamical effects observed in the sky \citep{2019ApJ...884...51G,2023arXiv230604837V}. In this case, $\langle v_{b}\rangle$ is about $12\pm4$ km/s in the first-infall scenario, lower by about 1 km/s compared to the blue shaded region, and decreases to below $7\pm4$ km/s in the second-passage scenario.

Moreover, the purple shaded region represents scenarios where the spherical MW halo transitions to an oblate shape ($q$ = 0.7), with its minor axis perpendicular to the Galactic disk. This change increases the transition MW halo mass between the first-infall and second-passage scenarios of the LMC's orbital history to $1.5\times10^{12} \mathrm{M}_{\odot}$ and reduces the magnitudes of $\langle v_{b}\rangle$ in both scenarios to lower values compared to previous cases. Specifically, $\langle v_{b}\rangle$ is about $12\pm4$ km/s in the first-infall scenario, where the MW halo mass ranges from $1.0\times10^{12} \mathrm{M}_{\odot}$ to $1.4\times10^{12} \mathrm{M}_{\odot}$, and about $4\pm4$ km/s in the second-passage scenario, with the halo mass ranging from $1.5\times10^{12} \mathrm{M}_{\odot}$ to $1.8\times10^{12} \mathrm{M}_{\odot}$.

As a result, our analysis of the kinematic properties of the RR Lyrae sample (yellow shaded region), which has a mean latitudinal velocity $\langle v_{b}\rangle=18.1\pm4.1$ km/s supports the first-infall scenario as the likely orbital history for the LMC. This value is consistent with a recent measurement from \cite{2024arXiv240601676C} using red giant stars. In particular, for the MW-LMC system where the LMC has a mass of $2.1\times10^{11} \mathrm{M}_{\odot}$ at infall, the LMC-induced perturbations in the first-infall branch (with $M_{\text{MW}}\leq 1.0\times10^{12}\mathrm{M}_{\odot}$) have mean latitudinal velocities of approximately $16\pm4$ km/s. This aligns more closely with observational data than scenarios involving a less massive LMC ($M_{\text{LMC}} = 1.5\times10^{11} \mathrm{M}_{\odot}$), where the observed mean latitudinal velocity is $\langle v_{b}\rangle\approx13\pm4$ km/s, regardless of the MW halo mass and velocity profile. Therefore, future studies on the MW-LMC interaction should consider LMC models with a virial mass close to $2.1\times10^{11}\mathrm{M}_{\odot}$ at the time of infall. Furthermore, in simulations with an oblate MW halo ($q$ = 0.7), both the first-infall and second-passage branches fall below the average latitudinal velocity in observation. This suggests that, for the simplified, non-tilted, spheroidal MW model aligned with the Galactic disk, the flattening parameter $q$ should be larger than 0.7. For example, if the MW halo has a flattening parameter of 0.8, it's plausible that the mean latitudinal velocities in the first-infall scenario (with $M_{\text{MW}}\leq1.3\times10^{12}\mathrm{M}_{\odot}$, for further discussion on the degeneracy between the transition MW mass and the flattening parameter, refer to section \ref{sec:mass-shape degeneracy}) can align with observational data when the LMC has a large virial mass at infall.
 
\section{Discussion}
\label{sec:discussion}
In this study, we aim to constrain the orbital history of the LMC by analyzing its dynamical effects on the Milky Way halo. We investigated the magnitude of perturbations in the velocity field of the MW's outer halo across various parameter combinations of the MW-LMC model and identified a discontinuity between the first-infall and second-passage scenarios. By comparing the latitudinal velocities of stars in our RR Lyrae sample with those derived from our simulated star catalogs, we found support for the first-infall scenario as the most probable orbital history for the LMC, while also placing constraints on the potential models of both galaxies. In the following, we will further evaluate the alignment of our findings with other observations, discuss some details of the MW-LMC simulations we have constructed, highlight the caveats and limitations of our study and explore potential directions for future research.

\subsection{Comparison to Previous Work}

On the LMC's mass, we find that the magnitudes of kinematic perturbations in both orbital scenarios are proportional to mass of the LMC and the magtitudes of the mean latitudinal velocity in the first-infall scenario exhibit a closer alignment with the observation of RR Lyrae stars from Gaia when the LMC mass is around $2.1 \times 10^{11} \mathrm{M}_{\odot}$ at its infall. Due to the tidal forces exerted by the MW, the current virial mass of the LMC is less than what it was before its entry into the MW. It is estimated that the LMC is likely to experience a reduction in mass by approximately 25$\%$ when it reaches the first pericenter of its orbit \citep{2023arXiv230604837V}. Thus, our constraint is consistent with some recent estimates of the mass of the LMC. For example, \cite{2016MNRAS.456L..54P} find that the LMC's total virial mass before its infall is most likely $M_{\mathrm{LMC}}=2.5_{-0.8}^{+0.9} \times 10^{11} \mathrm{M}_{\odot}$ by fitting the measured distances and velocities of galaxies in the
Local Volume as well as the relative motion between the MW and M31 derived from HST observations to a Bayesian model with the LMC mass as a free parameter. 

Moreover, the LMC's dynamical influence on stellar streams and kinematics of MW satellites are relevant to its total mass. \cite{2021ApJ...923..149S} fit the LMC mass (during infall) to be $1.88_{-0.40}^{+0.35} \times 10^{11} \mathrm{M}_{\odot}$ with five stellar streams with full 6D phase-space measurements in the proximity of the LMC. \cite{2022MNRAS.511.2610C} employ the likelihood-based forward modelling method and treat the LMC's mass as a model parameter, together with the unperturbed potential of the MW and the distribution function of the tracer in MW halo. They construct a time-dependent MW-LMC potential and rewind the orbits of globular clusters and satellite galaxies measured by Gaia backward to the time when the LMC has negligible influence to the MW. They then evaluate the likelihood of the tracers in their Bayesian model and the posterior of the LMC's mass (at infall) is $1.65_{-0.49}^{+0.47} \times 10^{11} \mathrm{M}_{\odot}$. These two results are broadly consistent with our measurement, which probes the LMC infall mass is around $2.1 \times 10^{11} \mathrm{M}_{\odot}$.

As for the shape of the MW halo. The majority of previous works report an oblate spheroid halo that is aligned with the Galactic disk and the flattening parameter $q$ is constrained to values close to 1.0 in some recent works. \cite{2015ApJ...803...80K} compared synthetic models of the Pal 5 stream's tidal tails with locally overdense regions of selected stars from SDSS DR9, finding that the halo of the MW has a flattening parameter of $q=0.95_{-0.12}^{+0.16}$. \cite{2018ApJ...859...31H} use 11,000 RRab stars (a subtype of RR Lyrae star) within $20 \mathrm{kpc} \leq R_{\mathrm{gc}} \leq 131 \mathrm{kpc}$ from Pan-STARRS1 to measure the density profile of the Galactic halo. The distribution of the sources is then well fit by a ellipsoidal stellar density model with a constant flattening parameter $q$ and $q=0.918_{-0.014}^{+0.016}$. In addition, \cite{2021MNRAS.508.5468H} model the stellar halo distribution function with the astrometric data for halo RR Lyrae stars within 30 kpc from Gaia
DR2 and find a posterior distribution of $q>0.963$. These measurements are consistent with our results obtained from the kinematic perturbations induced by the LMC in the MW halo, which favors $q > 0.7$. 

However, the shape of the real MW halo can be more complex than our simple model, which is characterized by a single flattening parameter $q$. For example, $q$ can be different in the inner and outer halo. \cite{2018ApJ...859...31H} fit their star sample with a ellipsoidal stellar density model with a radius-dependent flattening parameter and find the stellar halo is more flattened ($q \sim 0.8$) at inner halo (r $ \sim 25$ kpc) while more spherical  ($q \sim$ 1.0) at outer halo. Furthermore, MW halo may tilt relative to the stellar disk plane and present a triaxial shape. This has been demonstrated by fitting the phase-space structures of tidal streams spanning large arcs in the sky. Using the Sagittarius stream, \cite{2010ApJ...714..229L} and \cite{2021MNRAS.501.2279V} introduce a non-axisymmetric potential/density model, which features a nearly oblate ($q\sim0.7$) outer halo with its minor axis lying in the disc plane and pointing approximately in the direction of the LMC's orbital pole. Meanwhile, considering a different halo model with an axisymmetric shape and a radially constant flattening in the potential, \cite{2019MNRAS.487.2685E} also find that the direction of flattening is roughly aligned with the orbital plane of the LMC through the orbit-fitting of the Orphan stream. Beyond evidence from large-scale stellar streams, \cite{2022AJ....164..249H} use 5559 halo stars belonging to GSE within the radial range 6-60 kpc to infer the density profile of the accreted stellar halo. They find the shape of the accreted halo can be fit by a tilted, triaxial ellipsoid with the principal axes having a ratio of 10:8:7 and the major axis pointing 25 degree above the Galactic plane towards the position of the sun. Extending the analysis beyond our simple parametrized models in future studies will be crucial for investigating these subtle properties of the MW halo.

\subsection{Why the second-passage scenario induces less perturbation}
\label{sec: low_second_passage}

\begin{figure*}
	\includegraphics[width=2.0\columnwidth]{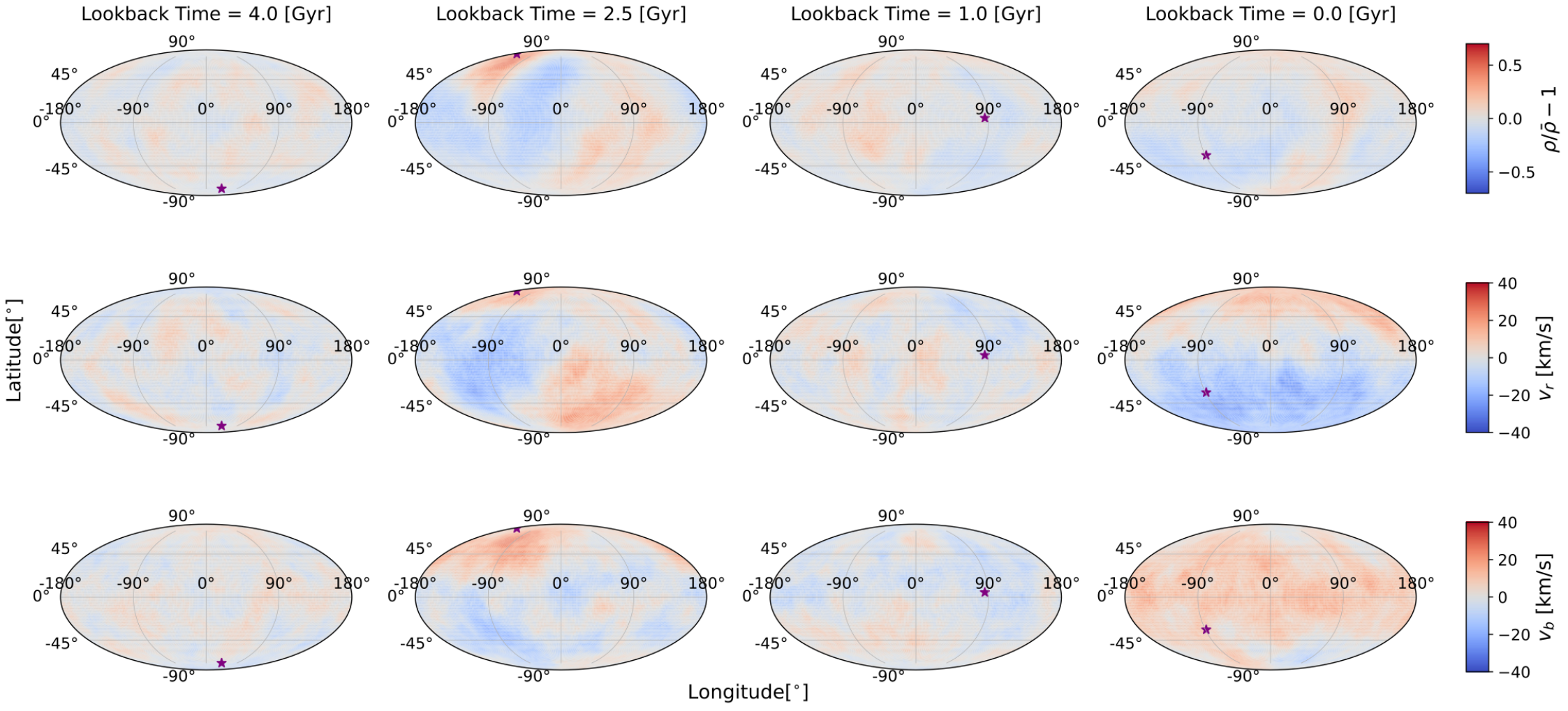}
    \caption{The time evolution of the local and global dynamical effects induced by the LMC in the second-passage scenario. We adopt the same MW-LMC system shown in Figure \ref{fig:1.3_map}, where $M_{\text{MW}}=1.3 \times 10^{12} \mathrm{M}_{\odot}$, $M_{\text{LMC}}=1.5 \times 10^{11} \mathrm{M}_{\odot}$ and $q=1.0$. We present the LMC-induced dynamic effects in density field, radial velocity ($v_{r}$) and latitudinal velocity ($v_{b}$) within the Galactocentric radius of 60-90 kpc. In these maps, the location of the LMC in the $l$-$b$ space is represented by a purple diamond. At the lookback time of 2.5 Gyr, the LMC had recently passed the first pericenter of its orbit, inducing perturbations in both density and velocity fields. When the LMC reached the apocenter position of its orbit and began its second infall about 1 Gyr ago, the inner MW established a new equilibrium with its surroundings and there is negligible relative displacement between the inner and outer halo. Thus, the local and global dynamical effects observed at present time are primarily a result of the LMC's second infall.}
    \label{fig:Time_map}
\end{figure*}

\begin{figure}
	\includegraphics[width=1.0\columnwidth]{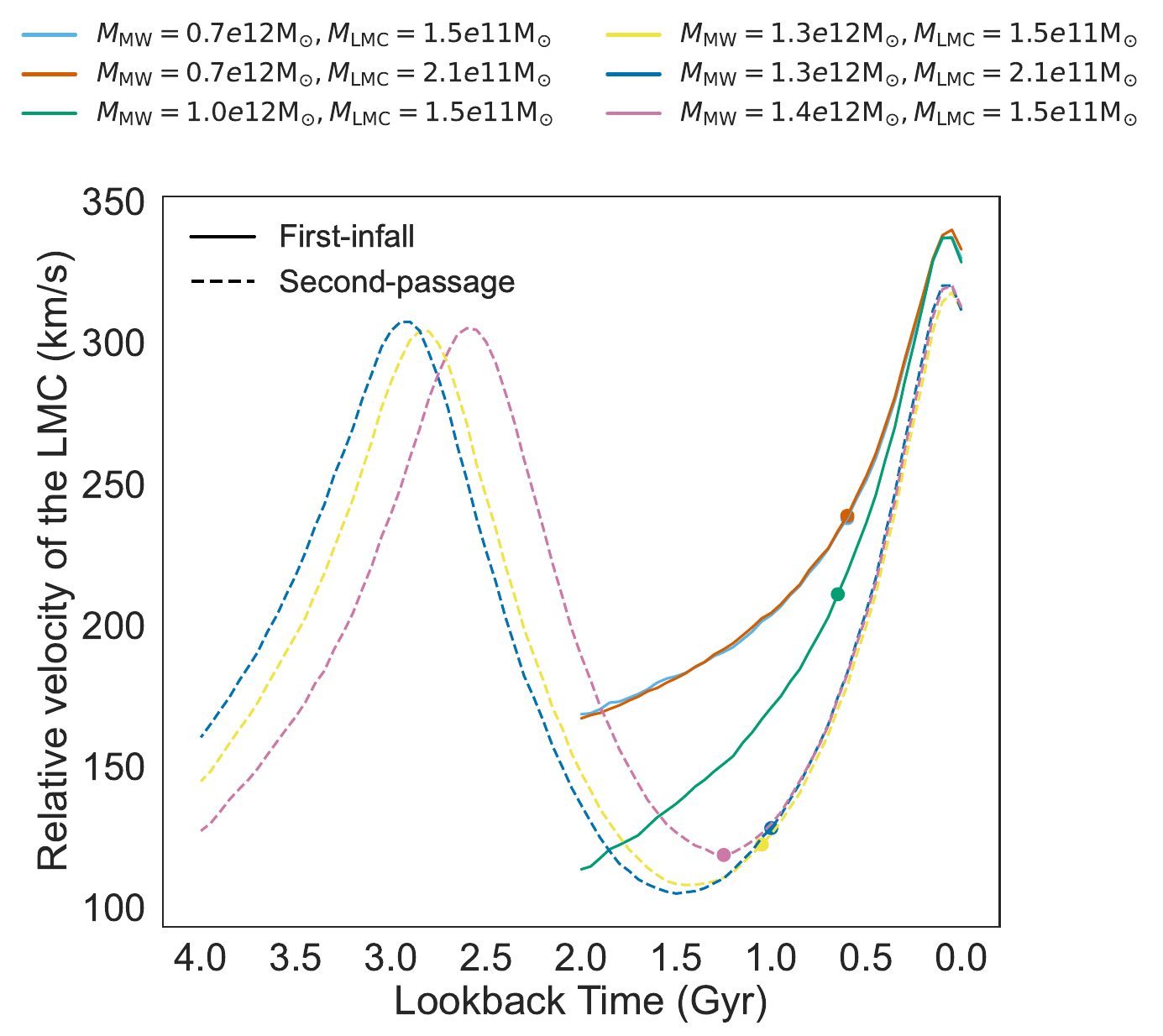}
    \caption{The evolution of the LMC's velocities relative to the Milky Way as a function of lookback time in both the first-infall and second-passage scenarios. The dot on each trajectory indicates when the LMC is approximately 140 kpc from the MW, marking the onset of its second infall. Due to the lower relative velocity of the LMC at the beginning of its second infall in the second-passage scenario, the Galactic inner halo experienced a reduced acceleration compared to the first-infall scenario. Consequently, this leads to a smaller displacement with respect to the outer halo and results in weaker global dynamical effects induced by the LMC.}
    \label{fig:velocity_compare}
\end{figure}

As demonstrated in earlier sections, the magnitude of perturbations in the MW halo is reduced in the second-passage scenario compared to the first-infall scenario owing to the short dynamical times of the Galactic inner halo. Therefore, the present-day perturbations in the second-passage scenario are predominantly due to the LMC's second infall into the MW, which starts at a much lower velocity relative to the inner halo than in the first-infall scenario.

To better understand why the second-passage generally leads to less perturbation, we show the time evolution of the local and global dynamical effects induced by the LMC in one representative second-passage scenario (the same MW-LMC system shown in Figure \ref{fig:1.3_map}, where $M_{\text{MW}}=1.3 \times 10^{12} \mathrm{M}_{\odot}$, $M_{\text{LMC}}=1.5 \times 10^{11} \mathrm{M}_{\odot}$ and $q=1.0$) in Figure \ref{fig:Time_map}. From left to right, each column represents the perturbations in density (top row), radial velocity (middle row) and latitudinal velocity (bottom row) fields of the MW halo within a Galactocentric distance range of 60-90 kpc at lookback times of 4.0, 2.5, 1.0, 0.0 Gyr, respectively. Throughout this time period, the LMC orbits the MW in a clockwise direction, with its position relative to the center of the MW marked by a purple diamond in the figure. 

The magnitudes of perturbations are reduced in the second-passage scenario due to the short dynamical time (less than 0.3 Gyr) of the Galactic inner halo (inner 30 kpc); at the lookback time of 2.5 Gyr, the LMC had recently passed the first pericenter of its orbit, inducing perturbations in both density and velocity fields. Importantly, when the LMC reached the apocenter position of its orbit and began its second infall about 1 Gyr ago, the inner MW established a new equilibrium with its surroundings and there are negligible relative displacements between the inner and outer halo induced by the LMC's first passage. Thus, the local and global dynamical effects observed at present time are primarily a result of the LMC's second infall. 

As the present-day dynamical effects observed in the second-passage scenario are largely induced by the LMC's most recent infall, it is crucial to figure out why the latter of the two passages results in a smaller perturbation compared to the first-infall scenario. Figure \ref{fig:velocity_compare} demonstrates the variations of the velocities of the LMC relative to the MW as a function of look-back time in some representative MW-LMC first-infall (solid line) and second-passage scenarios (dashed line). 
Both of the MW models have a spherical halo and a isotropic velocity profile. In the second-passage scenario, when the LMC begins its second infall at a lookback time of about 1 Gyr, its distance and velocity relative to the MW are approximately 140 kpc and 130 km/s, respectively. However, in the first-infall scenario, when the LMC reaches the same distance to the MW, the lookback time is about 0.6 Gyr, and at that time, the LMC's relative velocity exceeds 200 km/s. To maintain the same present-day velocities relative to the MW, the Galactic inner halo experiences a smaller acceleration in the second-passage scenario, resulting in a smaller displacement with respect to the outer halo and weaker global dynamical effects induced by the LMC. We also find that there is no significant difference in the relative velocity of the LMC at its second infall when its mass increases from $1.5 \times 10^{11} \mathrm{M}_{\odot}$ to $2.1 \times 10^{11} \mathrm{M}_{\odot}$. Therefore, even though the LMC could have lost approximately 25$\%$ of its mass during its first pericenter passage \citep{2023arXiv230604837V}, the diminished kinematic perturbations observed in the second-passage scenario should not be impacted by the mass loss of the LMC.

\subsection{The LMC impact has negligible on the net rotation of the halo.}

Given that a large fraction of stars in the MW's stellar halo are tidal remnants from dwarf galaxy mergers, the net rotation of the MW's stellar halo reflects the cumulative angular momentum from its historical merger events. In addition, the relationship between the net rotation of the stellar halo, the dark matter halo and the Galactic disk is fundamental to our understanding of the galaxy formation theory. If the massive LMC can induce kinematic perturbations in the MW halo, it is natural to wonder if such impact also generates visible net rotation of the MW halo.

Unlike the radial and latitudinal velocities, we find that the mean longitudinal velocity in the Galactocentric coordinate (shown in the bottom row of Figure \ref{fig:stat_mean}) remains close to zero and is unaffected by variations in different model parameters. Since all models of the MW in our simulations are built without halo net rotation ($\left\langle v_l\right\rangle= 0$), we find that the LMC's infall has a negligible influence on the net rotation of the MW stellar halo. This minimal impact is likely due to the LMC's orbital plane being nearly perpendicular to the Galactic disk plane, as demonstrated in Figure \ref{fig:velocity_plot}. Therefore, for any observed net rotation of the MW, such as the one noted in \cite{2017MNRAS.470.1259D}, where they find evidence for a rotating prograde signal $\left\langle v_l\right\rangle \sim 5-25$ km/s out to the Galactocentric radius of 50 kpc in RR lyrae, BHB and K giant stars), is likely a result from either the primordial Milky Way or the accumulation of the other merger systems.

\subsection{LMC's pericentric passage in 5-10 Gyr ago}
In some cases characterized as first-infall scenarios, the LMC could have made an additional pericentric passage 5-10 Gyr ago if its orbit was extended backward in time in our simulation. However, in this study, we only focus on second-passage scenarios where the LMC undergoes two pericentric passages within the last 5 Gyr. This choice is primarily due to the significant increase in the mass of the MW during the past 5-10 Gyr. In particular, the Milky Way has grown by approximately 50$\%$ of its current mass over the past 10 Gyr and reached about 80-90$\%$ of its present-day mass 5 Gyr ago \citep{2020MNRAS.497..747S}. However, in our dark matter-only N-body simulations, the MW potential is modeled as static over time, apart from the perturbations induced by the LMC. This modeling simplification introduces a discrepancy, suggesting that our simulated orbits of the LMC might be less reliable if extended backward in time to 5–10 Gyr ago, as the actual LMC would likely be on a less bound orbit due to the MW's mass growth. For example, \cite{2013ApJ...764..161K} demonstrates that the cosmological mass evolution of the MW can lead to a 20-30$\%$ increase in the LMC's orbital period. In addition, \cite{2023Galax..11...59V} shows that the LMC's orbit in a MW potential model which grow linearly with time is equivalent to a static MW model with 90$\%$ of its current mass.

Another reason for our choice to ignore the dynamical evolution beyond a lookback time for 5 Gyr has been inspired by \cite{2023arXiv230604837V}, where they showed that the second-passage scenario with the previous pericentric passage of the LMC occurring 5-10 Gyr ago, shows almost identical dynamical effects in the density and velocity fields of the MW halo compared to the first-infall scenario. This similarity is due to the fact that when the LMC underwent another pericentric passage 5–10 Gyr ago, its pericenter distance exceeded 100 kpc. Consequently, the LMC's apocenter distance and velocity at its second infall were large enough to induce indistinguishable perturbations from those in the first-infall scenario. This degeneracy might be overcame by a more detailed look into the star formation histories of the Magellanic Clouds \citep{2009AJ....138.1243H,2020A&A...639L...3R,2021MNRAS.508..245M,2022MNRAS.513L..40M}. In fact, there are two subdued star formation occured $\sim$ 5 and 8 Gyr ago \citep{2014MNRAS.438.1067M,2018MNRAS.473L..16M,2020A&A...639L...3R}, but no conclusive studies have proved their association with the LMC's previous approach to the Milky Way.

\subsection{Possible stellar halo sloshing from GSE}

\begin{figure}
	\includegraphics[width=1.0\columnwidth]{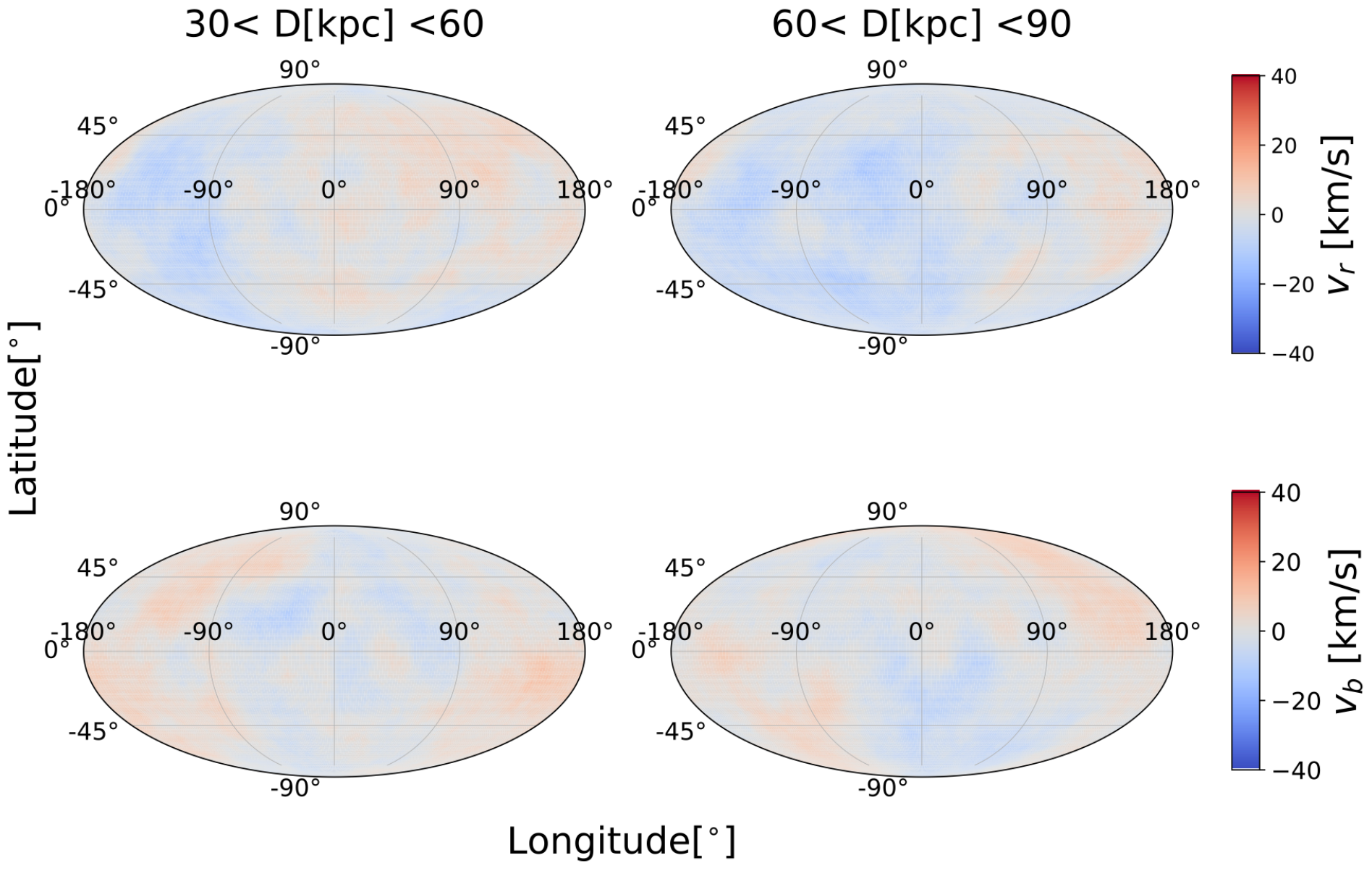}
    \caption{The ongoing sloshing of GSE on radial and latitudinal velocities within two radii intervals (30-60 kpc and 60-90 kpc) in the Galactocentric frame. Our analysis is conducted through a similar N-body simulation of the MW-GSE merger as described in \protect\cite{2021ApJ...923...92N}. We find that the remaining sloshing in both radial and latitudinal velocities shows average values of -0.86 km/s and -0.23 km/s for $30 \mathrm{kpc}<R_{\text{gal}}<60\mathrm{kpc}$, and -2.14 km/s and 0.34 km/s for $60 \mathrm{kpc}<R_{\text{gal}}<90\mathrm{kpc}$, respectively. The magnitudes of these perturbations are negligible compared to the predicted influences of the LMC if it follows a first-infall orbital scenario.}
    \label{fig:GSE_map}
\end{figure}

For stars in the outer halo of the MW, their dynamical times are long (e.g., for stars at a galactocentric distance of 100 kpc, their orbital periods are 1-2 Gyr) and they may need several dynamical times to reach equilibrium after the perturbation induced by previous significant mergers. For this reason, the last major merger of the MW, the Gaia-Sausage-Enceladus (GSE), could still have some residual sloshing in the MW's outer halo \citep{2021MNRAS.506.2677E}. 

To investigate whether this sloshing still exists, we run a similar N-body simulation of the MW-GSE merger as described in \cite{2021ApJ...923...92N}. They reconstruct the initial phase space coordinate of the GSE before it interacts with the MW by exploring a large grid of simulations spanning reasonable orbital and structural parameters to fit the observational constraints from H3 survey, including the spatial distribution of GSE debris, the existence of a highly retrograde debris (i.e. Arjuna), the emergence of Hercules-Aquila Cloud (HAC) and Virgo Overdensity (VOD)-like structures at the appropriate locations, the present-day stellar mass of GSE, the spatial extent of the MW's in-situ halo and the timing and duration of the MW-GSE merger. The model parameters of the MW and GSE describes their initial state before the merger at z $\sim$ 2. The initial MW model has a $5.0\times10^{11}\mathrm{M}_{\odot}$ halo, a $6.0\times10^{9}\mathrm{M}_{\odot}$ disk and a $1.4\times10^{10}\mathrm{M}_{\odot}$ bulge. The GSE model consists of a $2.0\times10^{11}\mathrm{M}_{\odot}$ halo and a $5.0\times10^{8}\mathrm{M}_{\odot}$ disk. The initial velocity of the GSE is determined so that its kinetic energy matches that of a circular orbit at the virial radius of the MW. This approach aligns with the behavior of infalling satellite galaxies observed in some cosmological dark matter simulations \citep{2015MNRAS.448.1674J,2017MNRAS.464.2882A}. The circularity $\eta$ and the orbital inclination $\theta$ are 0.5 and $15^{\circ}$ respectively and GSE entered the MW on a retrograde orbit. 

The results of the present-day simulation snapshot are shown in Figure \ref{fig:GSE_map}. Similar to Figure \ref{fig:0.7_map}-\ref{fig:1.3_map}, we illustrate the influence of GSE on radial and latitudinal velocities within two radii intervals (30-60 kpc and 60-90 kpc) in Galactocentric frame. We find that the remaining sloshing in both radial and latitudinal velocities shows average values of -0.86 km/s and -0.23 km/s for $30 \mathrm{kpc}<R_{\text{gal}}<60 \mathrm{kpc}$, and -2.14 km/s and 0.34 km/s for $60 \mathrm{kpc}<R_{\text{gal}}<90 \mathrm{kpc}$, respectively. Compared to the three representative cases demonstrated in Figures \ref{fig:0.7_map}-\ref{fig:1.3_map}, 
when the LMC follows a first-infall orbital scenario, the residual sloshing from the GSE, especially in the latitudinal velocity, is negligible compared to the predicted influences of the LMC. However, if the LMC is in the second-passage scenario, these residual effects from the GSE may be comparable on the LMC-induced perturbations in radial velocities. Given that our study supports the first-infall scenario for the LMC, the potential impact of stellar sloshing from the GSE on the LMC-induced kinematic perturbations is minimal.

\subsection{Mass-shape degeneracy}
\label{sec:mass-shape degeneracy}

Our findings suggest that the second-passage scenario is unlikely due to the low $\left\langle v_b\right\rangle$, irrespective of the MW halo's shape. As explored in Section \ref{sec: LMC orbital history}, the shape of the MW halo, defined by the flattening parameter $q$, affects the initial kinetic energy of the LMC. Moreover, the interplay between the MW halo shape and the virial mass determines whether the LMC undergoes a first-infall or a second-passage scenario. We delve deeper into the interplay between the MW halo's virial mass and its shape in the context of the LMC's orbital history, identifying which combinations of $q$ and $\mathrm{M_{MW}}$ result in the first-infall scenario, as contrasted with the second-passage scenario.

To simplify our analysis, we will fix the LMC virial mass at $1.5 \times10^{11}\mathrm{M}_{\odot}$. We also consider only the analytic orbit integration and integrate the LMC's orbit backward in time using galpy. The detailed initial conditions described in Section \ref{sec:orbital reconstruction} can be computationally expensive, and we found that while the details can slightly change the LMC's initial phase-space coordinates, they do not qualitatively change whether the orbit belongs to the first-infall or second-passage scenario. 

\begin{figure}
	\includegraphics[width=1.0\columnwidth]{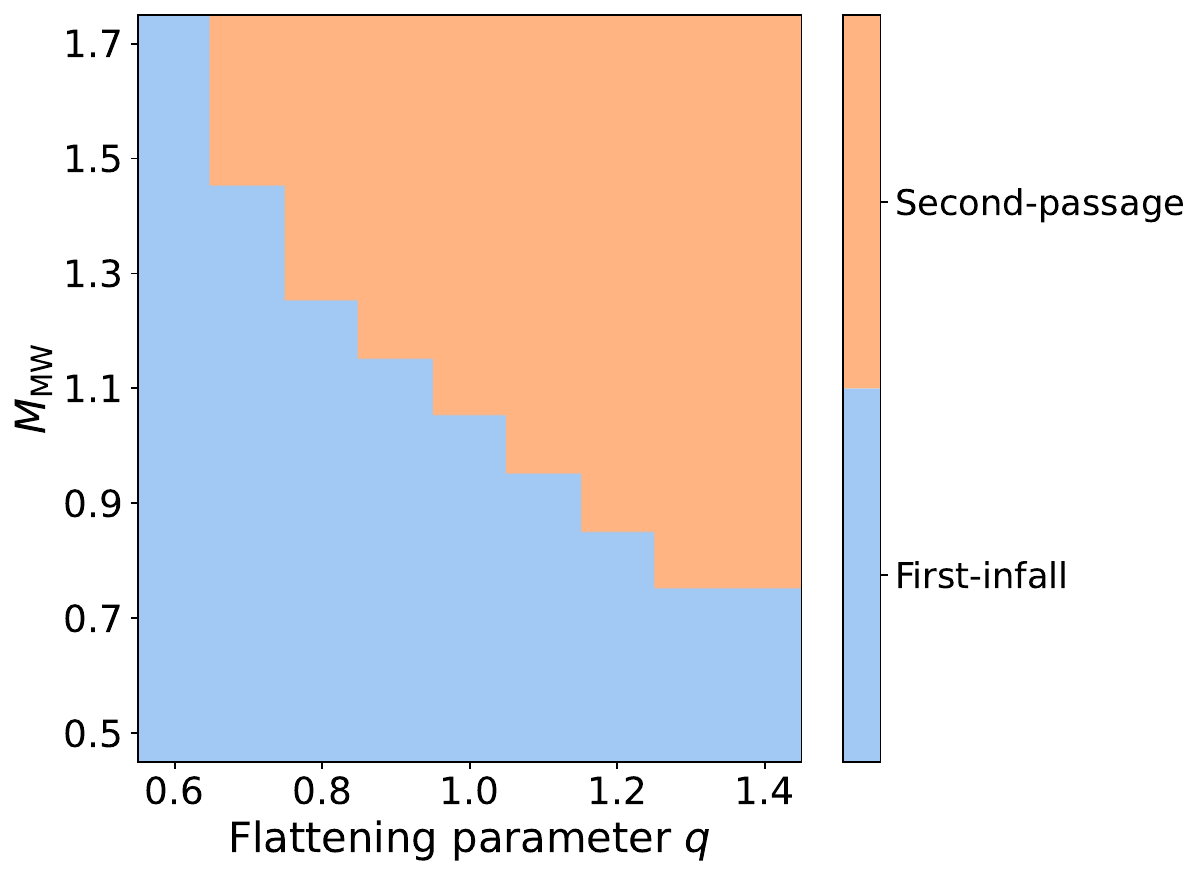}
    \caption{The degeneracy between the mass and shape (represented by the flattening parameter $q$) of the MW halo in determining the LMC's orbital history. For simplicity, the mass of the LMC is fixed at $1.5\times10^{11}\mathrm{M}_{\odot}$. To determine which orbital scenario fits these combinations of MW-LMC parameters, we integrate the LMC's orbit backward in time using galpy. When considering Milky Way halo models with a reduced virial mass and a higher degree of oblate shape (lower flattening parameter), the orbit of the LMC around it aligns with a first-infall orbital scenario. On the other hand, when the MW halo is more massive and exhibits a prolate shape, the LMC's orbital history follows a second-passage scenario. In particular, when the flattening parameter is within the range of 0.8-1.3, a 0.1 increase (decrease) in flattening parameter will lead the turnover mass to decrease (increase) $0.1\times10^{12}\mathrm{M}_{\odot}$.}
    \label{fig:mass_q}
\end{figure}

Figure \ref{fig:mass_q} shows the results. The $x$-axis represents the flattening parameter $q$, ranging from 0.6-1.4, while the $y$-axis represents the virial mass of the halo, ranging from $0.5-1.7\times10^{12}\mathrm{M}_{\odot}$. We find that when considering models of the MW halo with a lower virial mass and a more oblate shape (indicated by a lower flattening parameter), the LMC's orbit tends to correspond to a first-infall orbital scenario. Conversely, when the MW halo is more massive and exhibits a prolate shape, the LMC's orbital history is more consistent with a second-passage scenario. Specifically, when the flattening parameter lies within the range of 0.8 to 1.3, a 0.1 increase (or decrease) in the flattening parameter leads to a decrease (or increase) in the turnover mass by $0.1\times10^{12}\mathrm{M}_{\odot}$.






Drawing from our previous results regarding the shape of the MW halo, where the flattening parameter $q$ should be larger than 0.7, and considering that the LMC follows a first-infall orbital scenario, we can approximately constrain the halo mass of the MW to $\mathrm{M}_{\text{MW}}<1.4\times10^{12} \mathrm{M}_{\odot}$. This constraint is consistent with recent estimates in the literature that suggest the MW's virial mass is smaller than $1.5\times10^{12}\mathrm{M}_{\odot}$ \citep{2020SCPMA..6309801W,2022ApJ...925....1S}. The degeneracy between halo mass and shape that we explored in this section also highlights the importance for future studies to simultaneously constrain both of these parameters.

\subsection{Caveats and future directions}

This study aims to show that the second-passage scenario results in a perturbation too weak to align with observations. As our objective is not to make quantitative inferences about the parameters, we choose to make simplifying assumption on our N-body simulations, which we will elaborate on below.

The MW potential models in our simulations are built based on MWPotential2014 \citep{2015ApJS..216...29B}, while some other relevant studies use best-fitting MW parameters from \cite{2017MNRAS.465...76M}. In addition, our halo models generated from GalIC are spheroids aligned with the galactic disk and the only parameter we can adjust is the flattening parameter $q$. However, the real MW halo may tilt relative to the disk plane and present a triaxial shape due to the merger with GSE \citep{2022AJ....164..249H,2022ApJ...934...14H}.  

We also modelled the LMC as a single spherical halo and ignoring the gravitational influence from its most massive satellite, the SMC. However, although the current mass of the SMC is much less than that of the LMC ($M_{\text{SMC}}\sim 10^{10}\mathrm{M}_{\odot}$) and the SMC has negligible influence on the MW halo, some studies using numerical orbital integration have shown that it can still shift the LMC's trajectory. For example, \cite{2023arXiv230604837V} demonstrates that including a SMC with a mass one tenth of the LMC produces a non-negligible reduction in the LMC's orbital period and apocenter distance. In practice, simultaneously modelling the three-body interaction of the Milky Way, LMC and SMC in N-body simulations is challenging and largely an unresolved problem, particularly in determining the initial conditions of the Magellanic Clouds and matching their current position and velocity relative to the MW, which leads to our simplifying assumption in this study. Future work should consider the SMC's effect when studying the orbital history of the LMC.

In this study, we also primarily compare the mean latitudinal velocities of RR Lyrae stars in the Galactocentric frame with those from our simulated star catalogs, owing to the availability of 5-dimensional phase-space information for distant halo stars. However, our simulations suggest that (see Figure \ref{fig:stat_mean}) the radial velocities of halo stars in the southern hemisphere of the Milky Way can serve as an additional effective indicator of the LMC's orbital history. Unlike tangential velocities, radial velocities of halo stars can be measured with a higher degree of precision in ongoing and forthcoming spectroscopic surveys such as WEAVE \citep{2023MNRAS.tmp..715J}, 4MOST \citep{2016SPIE.9908E..1OD}, DESI \citep{2019BAAS...51g..57L} and PFS \citep{2022arXiv220614908G}. The availability of complete 6-dimensional phase-space information for these halo stars also enables us to investigate the perturbations in the Galactocentric frame, providing a robust view of the intrinsic density and kinematic properties of the MW halo.

Finally, although we characterize the magnitude of LMC-induce perturbations using simple summary statistics, the perturbed distribution of stellar density and kinematic features observed in simulations indicates that fitting the exact distribution of the phase space with the complete 6-dimensional phase-space information could lead to much stronger constraint on the MW-LMC model parameters from  the halo stars, provided sufficient simulations of the MW-LMC interaction are explored under various parameter combinations. This idea can be achieved through the modern simulation-based inference \citep[e.g.][]{2019MNRAS.488.4440A,2019ApJ...876....3L,2020PNAS..11730055C,2022ApJ...927..209T,2023ApJ...952L..10W}{}{}, which incorporates density estimation techniques based on neural networks to approximate the probability density (or likelihood) for the density and kinematic features of halo stars generated by our simulations or derived from direct observations. We will explore the feasibility of simulation-based inference in the MW-LMC interaction system in our coming work.

\section{Conclusions}
\label{sec:conclusion}

Understanding the orbital history of the Large Magellanic Cloud is essential for gaining insight into broader cosmological processes. The LMC's trajectory offers valuable perspectives on the formation and evolution of the Milky Way, as well as on the properties and distribution of dark matter at the galactic scale.

In this study, we construct a comprehensive grid of MW-LMC simulations by varying key model parameters (i.e. the virial mass and shape of the MW halo, the halo's velocity profile and the total mass of the LMC) of our fiducial model ($M_{\text{MW}}=1.0\times10^{12}\mathrm{M}_{\odot}$, $q=1.0$, $\beta(r)=0$, $M_{\text{LMC}}=1.5\times10^{11}\mathrm{M}_{\odot}$) and investigate how these parameters influence the magnitude of kinematic perturbations induced by the LMC. Additionally, we use high-resolution N-body simulations to illustrate the LMC-induced dynamical perturbations in the Milky Way halo in second-passage scenarios, where the LMC has made two pericentric passages within the last 5 Gyr. 
The key findings from this work can be summarized as follows:

\begin{itemize}
    \item We compare the mean latitudinal velocities in the Galactocentric frame of $\sim 1,100$ RR Lyrae stars located within a Galactocentric radius of 50-100 kpc with those from the mock star catalogs generated by our 50 MW-LMC simulations. We observe that the RR Lyrae sample exhibits a mean latitudinal velocity of $\langle v_{b}\rangle=18.1\pm4.1$ km/s. 
    \item This result aligns with the outcomes ($\langle v_{b}\rangle\approx16\pm4$ km/s) from simulations in which the LMC follows a first-infall orbital scenario and the MW halo is spherical ($q=1.0$). In contrast, the results from simulations where the LMC has a second-passage orbital scenario are consistently lower than $8\pm4$ km/s for various parameter combinations. Therefore, our comparative study supports the first-infall scenario for the LMC. 

    
    \item The reduced  magnitude of perturbation in the second-passage scenario is primarily attributed to the short dynamical timescales (e.g., 0.3 Gyr at the Galactocentric radius of 30 kpc) within the Milky Way's inner halo. As the LMC passes its apocenter and commences its second infall, the inner halo of the MW has already attained a new equilibrium with the outer halo. Therefore, the present-day global dynamical effects resulting from the MW's reflex motion are predominantly driven by the LMC's most recent (second) infall, where the LMC has a lower initial relative velocity and mass compared to those in the first-infall scenario.  

    \item The magnitude of kinematic perturbations is proportional to the mass of the LMC at infall when its orbital history is characterised by the first-infall orbital scenario. We find that for a spherical MW halo with an isotropic velocity profile ($\beta(r)=0$), a more massive LMC model with a mass of $2.1\times10^{11}\mathrm{M}_{\odot}$ induces stronger perturbation in the stars' latitudinal velocities in the first-infall scenario compared to a lighter LMC with a mass of $1.5\times10^{11}\mathrm{M}_{\odot}$, and is more consistent with the perturbations derived from the RR Lyrae sample.
    
    \item In simulations where the MW halo has an oblate shape with the flattening parameter $q= 0.7$, the perturbations in stars' latitudinal velocities in both first-infall and second-passage scenarios fall below the average latitudinal velocity in observation. This suggests that, for the non-tilted, spheroidal MW model aligned with the Galactic disk, its flattening parameter $q$ should be larger than 0.7. 
    
    \item With the flattening parameter $q>0.7$, only a MW virial mass lower than $1.4 \times 10^{12} \mathrm{M}_{\odot}$ satisfies the first-infall scenario of the LMC's orbital history, indirectly constraining the MW virial mass to be less than $1.4 \times 10^{12} \mathrm{M}_{\odot}$.

\end{itemize}

Our results demonstrate that the dynamical influence of the LMC and MW can lead to meaningful constraints on their properties. This study paves the way for ongoing and upcoming spectroscopic surveys such as WEAVE and 4MOST, which, in conjunction with more accurate proper motion data from Gaia, will provide us with complete 6-dimensional phase space information for more distant halo stars. These data will enable us to quantify the spatial distribution of LMC-induced perturbations in both density and velocity fields, allowing for a better understanding of the closest cosmic dance of the galaxies in our Galactic backyard.


\textit{Software}: IPython \citep{4160251}, matplotlib \citep{4160265}, seaborn \citep{Waskom2021}, numpy \citep{harris2020array}, scipy \citep{2020SciPy-NMeth}, jupyter \citep{Kluyver2016jupyter}, Astropy \citep{astropy:2013,astropy:2018}, Gadget-4 \citep{2021MNRAS.506.2871S}, GALIC \citep{2014MNRAS.444...62Y}, galpy \citep{2015ApJS..216...29B}, h5py \citep{collette_python_hdf5_2014}, gala \citep{2017JOSS....2..388P}.

\section*{Acknowledgements}
We are grateful to Nico Garavito-Camargo and Eugene Vasiliev for advice on simulations. Y.S.T. acknowledges financial support from
the Australian Research Council through DECRA Fellowship DE220101520. X.-X.X. acknowledge the support from National Key Research and Development Program of China No. 2019YFA0405504, National Natural Science Foundation of China (NSFC) under grants No. 11988101, CAS Project for Young Scientists in Basic Research grant No. YSBR-062, and China Manned Space Project with No. CMS-CSST-2021-B03. J.C. acknowledges the support from the National Natural Science Foundation of China under grant No. 12273027. H.T. is supported by National Natural Science Foundation of China with grant No. 12103062 and the science research grants from the China Manned Space Project. We further acknowledge the high performance computing resources provided by the Australian National Computational Infrastructure (grants y89) through the National and ANU Computational Merit Allocation Schemes.

This work has made use of data from the European Space Agency (ESA) mission Gaia (https://www.cosmos.esa.int/gaia), processed by the Gaia Data Processing and Analysis Consortium (DPAC, https://www.cosmos.esa.int/web/gaia/dpac/consortium).

\section*{DATA AVAILABILITY}
The data underlying this article will be shared on reasonable request with the corresponding author.




\bibliographystyle{mnras}
\bibliography{example} 




\appendix
\section{Determine the orbit initial condition of the LMC with Gauss-Newton iteration}
\label{sec: Gauss-Newton}

Our method for reconstructing the LMC's orbits involves a constraint to match its present-day center-of-mass position and velocity in simulations with the measurements obtained from recent observations \citep{2013ApJ...764..161K}. Unfortunately, obtaining an appropriate initial phase-space coordinate of the LMC in N-body simulations by simply integrating the orbit backward in a static analytic MW potential model proves challenging due to the LMC-induced deformations of the MW and the our estimation of the dynamical friction. While we are not able to directly resolve these two issues, we can use the outcome of the previous numerical integration as a first guess for the LMC's initial condition and then iteratively refine this estimate in the N-body simulation where both the MW and the LMC models are represented as particle systems. Our refinement of the LMC's initial conditions is based on minimizing the discrepancies between the present-day coordinates of its simulated orbit and the observational values until they are sufficiently close. Following the methodology outlined in \cite{2020MNRAS.497.4162V} and \cite{2023arXiv230604837V}, we employ the Gauss–Newton iteration to find the orbital initial condition  $\boldsymbol{w}_{\mathrm{in}} \equiv\{x,v\}_{in}$ leading to the given final position and velocity $\boldsymbol{w}_{\text {fin}}$ after a fixed time $T_{\text{end}}$. 

The Gauss-Newton iteration is a numerical optimization technique primarily used for solving non-linear least squares problems. In our case, we have the actual observations $\boldsymbol{w}_{\text {fin, true }}$ and a non-linear model that depends on model parameters ( i.e. the initial phase-space coordinate of the LMC $\boldsymbol{w}_{\text {in }}$). The goal is to find the best set of parameters that minimizes the difference between the model's predictions $\boldsymbol{w}_{\text {fin}}$ and the actual observations. For a given first guess $\boldsymbol{w}_{\text {in }, 0}$, we follow an ensemble of initial condition with slightly offset $\boldsymbol{w}_{\text {in }, k}, k=1 . . K$ (we set K equals 6 in this work), which then produce K different final phase-space coordinate $\boldsymbol{w}_{\text {fin}, k}$ after evolving the same $T_{\text{end}}$ in the simulation. At each iteration, the Gauss-Newton method approximates the non-linear model by a linear one, based on the K+1 initial and final coordinates. The next choice of initial coordinate is given by $\boldsymbol{w}_{\mathrm{in}, 0}^{(\text {new })}=\boldsymbol{w}_{\mathrm{in}, 0}-\mathrm{J}^{-1}\left(\boldsymbol{w}_{\mathrm{fin}, 0}-\boldsymbol{w}_{\text {fin, true }}\right)$. The $\boldsymbol{w}_{\text {fin, true }}$ is the present-day phase space coordinate of the LMC that we used to constrain the orbit in simulation, while $\boldsymbol{w}_{\text {fin, 0 }}$ is the final state of our first guess. The Jacobian matrix $\mathrm{J} \equiv \partial \boldsymbol{w}_{\mathrm{fin}} / \partial \boldsymbol{w}_{\mathrm{in}}$ represents how the residual (difference between observed final coordinate and the final coordinate predicted by the model) change with small offsets in the initial coordinate, and is approximated by finite differences: $\mathrm{J} \approx \delta \mathrm{w}_{\mathrm{fin}} \delta \mathrm{w}_{\mathrm{in}}^{-1}$, where the columns of matrices $\delta \mathrm{w}_{\ldots}$ contain the difference vectors $\boldsymbol{w}_{\ldots, k}-\boldsymbol{w}_{\ldots, 0}$. We repeat this process until the residual is within $2\sigma$ of the observed final coordinate.

However, there are scenarios where the Jacobian matrix may become ill-conditioned, indicated by one or more singular values approaching zero. This tends to occur, especially when simulating over extended time, such as 4-5 Gyr. The underlying reasons for this may vary (e.g., the discreteness noise in N-body simulations), but in practice, we replace large singular values (larger than 10 in our case) in the inverse Jacobian with zeroes, and recompute the predicted next initial condition of the LMC with this manually tweaked Jacobian: $\boldsymbol{w}_{\mathrm{in}, 0}^{(\text {new })}=\boldsymbol{w}_{\mathrm{in}, 0}-V^{T}S_{\text{replace}}^{-1}U^{T}\left(\boldsymbol{w}_{\mathrm{fin}, 0}-\boldsymbol{w}_{\text {fin, true }}\right)$, where $J=USV$. This is a common technique in poorly or ill-conditioned least-square problems solved with the singular-value decomposition approach, also known as Moore-Penrose pseudo inverse. 

Typically, 5-10 iterations are sufficient for the derived phase-space coordinates of the LMC to converge to the observed values within $2\sigma$ in the first-infall scenario, with the complete set of iterations consuming approximately 800 CPU hours. In contrast, for the second-passage scenario, 10-20 iterations are necessary to achieve convergence, and the total iterations demand around 6,500 CPU hours.

\section{Convergence test of the simulations}
\label{sec: Convergence test}
\begin{figure}
	\includegraphics[width=1.0\columnwidth]{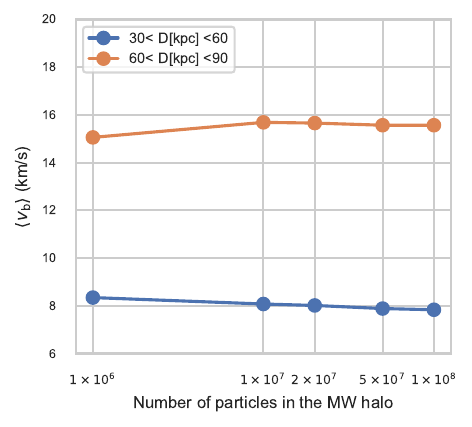}
    \caption{Convergence test of the number of particles used in our simulations. We make this test using the fiducial MW-LMC model ($M_{\text{MW}}=1.0 \times 10^{12} \mathrm{M}_{\odot}$, $M_{\text{LMC}}=1.5 \times 10^{11} \mathrm{M}_{\odot}$, $q=1.0$ and the halo has a isotropic velocity profile $\beta(r)=0$) with five different resolutions where the number of particles in the Galactic halo is $10^8$, $5\times10^7$, $2\times10^7$, $10^7$ and $10^6$, respectively. We show the mean latitudinal velocities within two Galactocentric distance ranges ($30<R_{\text{gal}}<60$ kpc $\&$ $60<R_{\text{gal}}<90$ kpc). Our findings indicate that the mean values remain stable in N-body simulations with a resolution of $10^6$ particles or greater.}
    \label{fig:convergence_map}
\end{figure}

In this section, we evaluate whether the number of particles used in our simulations is sufficient to capture the dynamical effects induced by the LMC and whether the mean values of velocity perturbations converge at our chosen resolution ($10^{8}$ particles). While previous studies such as \cite{2009MNRAS.400.1247C} and \cite{2019ApJ...884...51G} have suggested that a resolution of $10^{6}$ particles (with equal particle mass) is sufficient to capture the LMC-induced dynamical effects in the MW halo, and a resolution of $10^{8}$ is necessary to resolve smaller-scale structures, we investigate their results in the context of the mean values of velocity perturbations using our fiducial MW-LMC model ($M_{\text{MW}}=1.0 \times 10^{12} \mathrm{M}_{\odot}$, $M_{\text{LMC}}=1.5 \times 10^{11} \mathrm{M}_{\odot}$, $q=1.0$, and an isotropic velocity profile $\beta(r)=0$). Figure \ref{fig:convergence_map} shows the mean value of latitudinal velocities within two Galactocentric distance ranges ($30<R_{\text{gal}}<60$ kpc $\&$ $60<R_{\text{gal}}<90$ kpc) for five simulation resolutions where the number of particles in the Galactic halo is $10^8$, $5\times10^7$, $2\times10^7$, $10^7$ and $10^6$ respectively. Our results demonstrate that the mean values are stable to within 3$\%$ for the 30-60 kpc distance range and stable to within 4$\%$ for the 60-90 kpc distance range when the resolution of the N-body simulation is equal to or greater than $10^6$ particles.

\bsp	
\label{lastpage}
\end{document}